\documentclass{aa}

\usepackage{natbib}
\usepackage{graphicx}
\usepackage{txfonts}
\usepackage{color}

\newcommand{\ergs}{\rm\,erg\,s^{-1}}
\newcommand{\msun}{$M_{\odot}$}
\newcommand{\psrb}{PSR~B1257+12}
\newcommand{\mearth}{$M_{\oplus}$}
\newcommand{\fortyK}{$^{40}$K}

\newcommand{\eightU}{$^{238}$U}
\newcommand{\twoTh}{$^{232}$Th}

\begin{document}

\title{Neutron Star Planets: Atmospheric processes and habitability}
%  \thanks{AP acknowledges support from an NWO Vidi fellowship.}}

\subtitle{}

\author{A. Patruno\inst{1,2} \and M. Kama\inst{3,1}}

  \institute{
    Leiden
    Observatory, Leiden University, Neils Bohrweg 2, 2333 CA, Leiden, The Netherlands
    \and 
    ASTRON, the Netherlands Institute for Radio Astronomy, Postbus 2, 7900 AA, Dwingeloo, the Netherlands 
    \and
    Institute of Astronomy, University of Cambridge, Madingley Road, Cambridge, UK, CB3 0HA
    }

\date{}

\abstract{Of the roughly 3000 neutron stars known, only a handful have sub-stellar companions. The most famous of these are the low-mass planets around the millisecond pulsar B1257+12. New evidence indicates that observational biases could still hide a wide variety of planetary systems around most neutron stars. We consider the environment and physical processes relevant to neutron star planets, in particular the effect of X-ray irradiation and the relativistic pulsar wind on the planetary atmosphere. We discuss the survival time of planet atmospheres and the planetary surface conditions around different classes of neutron stars, and define a neutron star habitable zone. Depending on as-yet poorly constrained aspects of the pulsar wind, both Super-Earths around B1257+12 could lie within its habitable zone.
 }

\maketitle

\section{Introduction}

% Formation of neutron stars and their disks
Neutron stars are created in supernova explosions and begin their lives surrounded by a fallback disk with a mass of order $0.1$ to $0.2$~\msun~\citep{FryerHeger2000}, giving a disk-to-primary mass ratio similar to protoplanetary disks close to gravitational instability. These disks have a metallicity equal to or above the Galactic average, and are thus dust-rich. The
presence of a dusty disk with $10\,$\mearth\ of material has been
proposed to explain a mid-infrared excess around a young isolated
neutron star~\citep{wan06} and a mid-IR counterpart has been also
detected around the magnetar 1E 2259+286~\citep{kap09}. These discoveries are consistent with dusty fallback disks around neutron stars $\sim1\,$Myr after formation.

% Types of neutron stars
The neutron star family broadly consists of four categories,
according to the main mechanism that powers their emission:
\begin{itemize}
\item young radio pulsars (young PSRs, about $\approx2200$ objects known), powered by their rotational energy;
\item millisecond radio pulsars (MSPs, $\approx400$ objects), which have accreted matter from a companion star; 
\item thermally emitting dim isolated neutron stars (DINSs, 7 objects
  known), powered by their thermal cooling and accreting from the
  interstellar medium.
  \item accreting neutron stars (ANSs, a few hundred known objects),
    powered by accretion of gas from a companion star.
\end{itemize}
These comprise nearly all known neutron stars, with rare exceptions
like rotating radio transients, magnetars and central compact objects.
These distinctions are important because different physical processes
shape the environment around different neutron stars, leading to a
range of effects on their planets.

% Neutron star planets
The first exoplanets discovered were the three low-mass objects found around the
millisecond radio pulsar B1257+12 \citep{wol92}. One of these is a tiny object with
${\sim}0.02\,M_{\oplus}$ whereas the other two are Super-Earths of $\approx4\,M_{\oplus}$ \citep{wol94}. 
After B1257+12, other neutron stars have been found to host sub-stellar companions.
The ``diamond-planet'' system PSR J1719--1438 is a
millisecond pulsar surrounded by a Jupiter-sized companion thought to
have formed via ablation of its donor star~\citep{bai11}. The system
PSR B1620-26 instead is a millisecond pulsar-white dwarf binary
surrounded by a Jupiter-sized planet in a 40 years
orbit~\citep{tho93}. The latter system is in the globular cluster NGC
6121 (M4) and its formation is probably related to dynamical
interactions that occurred in the cluster~\citep{sig03}. It has also been
recently proposed that stochastic timing variations observed around
the millisecond pulsar PSR J1937+21 might be related to the presence
of an asteroid belt~\citep{sha13}, and a similar idea has been
suggested for the~\textit{young} PSR J0738--4042~\citep{bro14}. There
is even a gamma ray burst (GRB 101225A) which has been proposed to be
the consequence of a tidal disruption of a minor body around a neutron
star~\citep{cam11} and some authors have explained
the enigmatic phenomenon of fast radio bursts as asteroid/comet
collisions with neutron stars~\citep{gen15}.

%Why only few planets around PSRs? 
Neutron star planets can be first-, second- or third-generation. First
generation planets would be formed in the usual manner, as a
by-product of the star formation process, and would likely be ablated
or unbound during stellar death. Second generation objects would form
in the supernova fallback disk around a freshly-formed neutron
star. Third generation planets would form from a disk consisting of a
disrupted binary companion (possibly previously overflowing its Roche lobe),
thought to be essential for producing millisecond pulsars such as
B1257+12. The supernova explosion, the accretion from a companion for
millions up to billion years that MSPs undergo, and the emission of
high energy X-ray/$\gamma$--ray radiation and MeV--TeV particles (the
pulsar wind) are all disruptive processes that might destroy planets
or disrupt their orbits. Although, based on the fact that very few
planets have been found to date, \citet{mar16} have recently suggested
that the formation of planets around pulsars is an inherently rare
phenomenon, it is clear that planet formation can happen around neutron
stars. Furthermore, existing timing measurements of MSPs allow for
low-mass planets on larger orbits, and the presence of planets around
most non-MSP neutron stars is still essentially unconstrained.

% Can pulsar planets harbour life ? 
Even if the fraction of neutron star systems that form
planets is as small as the current detection fraction, the large number of neutron stars in the galaxy
\citep[$\sim10^{9}$;][]{kog92} still guarantees that a relatively
large number, perhaps $\gtrsim10^{7}$, of such planetary systems exist. 
We move one step forward and ask how different pulsar planets are
with respect to those found around main sequence stars,
whether they can retain an atmosphere, and whether they may be habitable.

Planetary systems around neutron stars need not be 
similar to the thousands planets that have been found around
main sequence stars \citep[e.g.][]{bor10,bat13,bur14}. For example, the habitability of a
planet is generally defined in terms of equilibrium surface
temperature set by the radiant energy it receives from its host
star. Such stellar radiation is in first approximation blackbody
radiation peaking at near-IR, optical or UV and typical habitable
zones are set at a fraction of, and up to a few, astronomical units.
A habitable zone of a drastically smaller size than around main sequence stars was derived
for white dwarfs by \citet{ago11}. They found that white dwarfs with a surface
temperature of less than $\approx10^{4}\,$K may host a
habitable planet for $\approx3\,$Gyr if it lies in a narrow band
$\sim0.005-0.02\,$au from the white dwarf.  In the case of neutron stars the
blackbody radiation emitted might peak in X-rays, copious amounts of
ionizing high energy particles might be present and almost no
near-IR/optical/UV radiation is emitted and therefore different
mechanisms need to be considered.

In this paper we address two problems related to neutron star
planetary systems. The first is to verify whether there is any further
evidence for the presence of a cloud of debris/gas around the planet-hosting \psrb, as
suggested by \citet{pav07}. In Section~\ref{sec:obs} and~\ref{sec:ana}
we thus present our analysis on archival 2007 \emph{Chandra}
observations of this system and discuss the limits we can set on the
presence of a cloud of absorbing material. The second goal is to
verify whether neutron star planets can harbour an atmosphere and what
are the conditions on their surfaces, which are bathed in
ionizing radiation and energetic particles.

We discuss the physical
conditions found in the environment around neutron stars in
Section~\ref{sec:env}.  We then discuss in Section~\ref{sec:irr} how
such harsh conditions can affect a planetary atmosphere.  Then we
apply our calculations to the observed population of pulsar planets
(Section~\ref{sec:pla}) and infer in which conditions they are likely
to harbor an atmosphere, which timescales are required for their
evaporation and finally we touch the question of a neutron star 
habitable zone. In Section~\ref{sec:comp}
we discuss potential differences in composition of neutron star planets,
related to the different type of environment in which they are formed.

\section{X-ray observations}\label{sec:obs}
\begin{figure}
\includegraphics[width=1.0\columnwidth]{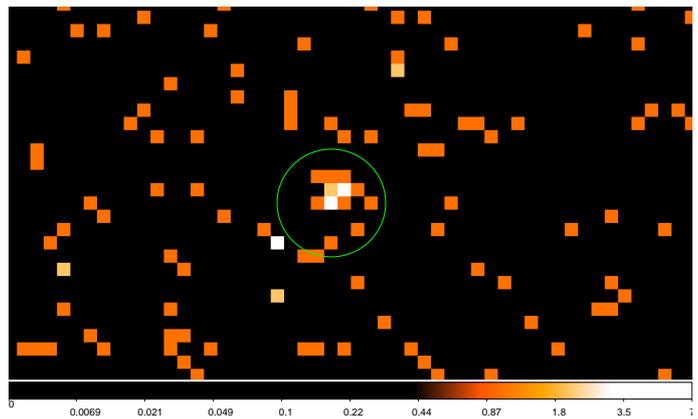}
  \caption{\psrb\, as seen by \textit{Chandra} on 2007, May 3. The
    green circle is the $1.5''$ radius extraction region used to find the source counts. The region is centered around the white pixel, which is the brightest one in the image. The color scale refers to number of counts.}
  \label{fig:src}
\end{figure}

We used archival \textit{Chandra} observations (ObsID: 7577) carried
on 2007 May 3 (start time 01:03:41 UTC) for a total of 18.31 ks. The
data were collected with the Advanced CCD Imaging Spectrometer (ACIS)
in Very Faint Mode.  The data were processed using the CIAO software
(v.4.9) along with the calibration database (CALDB; v.4.7.3). The
X-ray photons were identified with the SAOImage ds9 tool (v7.4) by
selecting a circular source region of $1.5''$ in radius centered
around the brightest pixel close to the coordinates of the pulsar
(RA:$13^h\,00^{m}\,03.0810^s$, DEC:$12^{\circ}\,40'55.875''$; see Figure~\ref{fig:src}).
The background was obtained from an annular region centered around the
source with inner radius of $3''$ and outer radius of $30''$. The
photons were then extracted with the tool \emph{specextr} which
calculates also response and ancillary files for X-ray spectral
analysis. The X-ray spectra were analyzed with the XSPEC software
(v12.9.1). We also use the archival \textit{Chandra} observation
(ObsID 5518) taken on May 22, 2005 (20.05 ks long) to reproduce the
results discussed in \citet{pav07}.

\section{Data analysis \& results}\label{sec:ana}

We begin with the 2005 observations to try to reproduce the results of
\citet{pav07}. We find 25 photons in the 0.3--8 keV band and we fit
the spectrum by using two models, a blackbody with normalized area
(\emph{bbodyrad} in XSPEC) and a simple power law. Both models are
fitted with absorption where we keep the hydrogen column density
$N_{\rm H}$ fixed at a value of $3\times10^{20}\rm\,cm^{-2}$ as
inferred from the pulsar timing (via the dispersion measure). We
find best fit parameters compatible within the 1$\sigma$ error with
those reported by the authors.

The 2007 observation shows a total of 21 source photons in the
0.3--8 keV energy range.  Given the small number of photons, we do not
group our data and use the C-statistics~\citep{cas79}.
The results of our spectral fit are reported in Table~\ref{tab:fit}.
\begin{table}[!ht]
\centering
\caption{X-Ray Spectral Models for the 2007 \textit{Chandra} Data}
\begin{tabular}{lccccc}
\hline\hline
Model & $N_{\rm H}$ & Normalization & $\Gamma$ or $kT$ & C/dof&\\
\hline\\
PL & $3$ & $1.16^{+0.41}_{-0.35}$ & $3.43^{+0.65}_{-0.62}$ & 65/524&\\\\
BB & $3$ & $1.15^{+1.85}_{-0.70}$ & $0.149^{+0.029}_{-0.023}$ & 65/524&\\
\\
\hline
\hline
\end{tabular}
\emph{\footnotesize{Note: The errors correspond to 68\% confidence intervals.  $N_{\rm H}$ is kept frozen during the fit and is given in units of $10^{20}\rm\,cm^{-2}$.  The power-law normalization is in units of $10^{-6}\rm\,erg\,cm^{-2}\,s^{-1}$ at 1 keV.}}
\label{tab:fit}
\end{table}

\begin{figure}
  \rotatebox{-90}{\includegraphics[width=0.75\columnwidth]{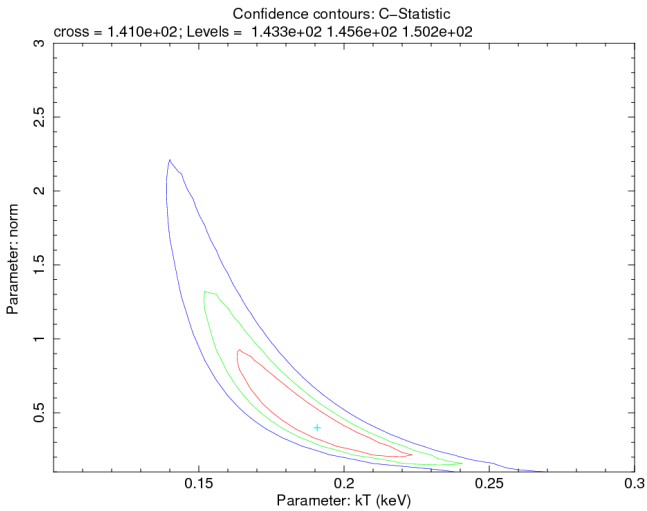}}
  \caption{Contour plot for temperature (units of keV)
    vs. normalization $K$. The contours refer to the 68\% (red line),
    90\% (green line) and 99\% (blue line) confidence
    levels. Degeneracy between temperature and normalization is
    observed, but it is still not sufficient to reconcile the small
    emitting area with theoretical models of millisecond pulsars.}
  \label{fig:cont1}
\end{figure}

Both a power-law and a blackbody model give results which are
compatible, within the statistical uncertainties, to those obtained for the
2005 \textit{Chandra} data. The 0.3--8.0 keV luminosity of the
power-law model corresponds to an X-ray luminosity of
$3.1^{+2.8}_{-1.3}\times10^{29}\rm\,erg\,s^{-1}$ for an assumed source
distance of 700 pc, whereas the blackbody model gives a very small
emission area of the order of $2300\rm\,m^2$, analogous to the
previous findings of 2005\footnote{the blackbody normalization $K$ is related
  to the projected area of the blackbody $A_{\perp,a}$ via the
  relation: $A_{\perp,a}=K\pi\,d_{10}^2\cdot\,10^6\rm\,m^2$, where
  $d_{10}$ is the distance in units of 10 kpc.}. Since the projected
area is very small, we explore the degeneracy of the blackbody
temperature and its normalization $K$ by plotting the confidence
contours for the 68\%, 90\% and 99\% level (see
Figure~\ref{fig:cont1}).  Even when using the most extreme values for
the normalization (at the boundary of the 99\% confidence contour,
$K\approx2.2$) the size of the emitting hot spot is
$A_{\perp,a}\approx 3\times10^4\rm\,m^2$, more than two orders of magnitude
smaller than the predicted theoretical values of $\sim10^7\rm\,m^2$.

Since we see no difference in the spectral parameters between the 2005 and
2007 observations, we try to fit the data simultaneously to increase the
signal-to-noise, which is valid under the assumption that the
underlying spectral shape has remained truly the same in the two observations. 
The results of our fit are reported in Table~\ref{tab:comb}.

\begin{table}[!ht]
\centering
\caption{X-Ray Spectral Model for the 2005 \& 2007 \emph{Chandra} Data}
\begin{tabular}{lllll}
\hline\hline
Model & $N_{\rm H}$ & Normalization & $\Gamma$ or $kT$ & C/dof\\
\hline
PL & 3 & $1.82^{+0.27}_{-0.24}$ & $2.93^{+0.31}_{-0.31}$ &145.5/1050\\\\
BB & 3 & $0.41^{+0.28}_{-0.17}$ & $0.19^{+0.022}_{-0.019}$ & 141/1050\\
\hline
NSA& 3 & <3.8 & $0.098^{+0.047}_{-0.020}$ & 140.8/1050\\
\hline\hline
\end{tabular}
\footnotesize{\emph{Note: The errors correspond to $68\,$\% confidence intervals. $N_{\rm H}$ is kept frozen during the fit and is given in units of $10^{20}\rm\,cm^{-2}$. The power-law normalization is in units of $\rm\,erg\,cm^{-2}\,s^{-1}$ at 1 keV. The neutron star atmosphere model normalization is in units of $10^{6}\,$m$^{2}$ and refers to a $90\,$\% confidence level upper limit.}}
\label{tab:comb}
\end{table}

\begin{figure}
\rotatebox{-90}{\includegraphics[width=0.75\columnwidth]{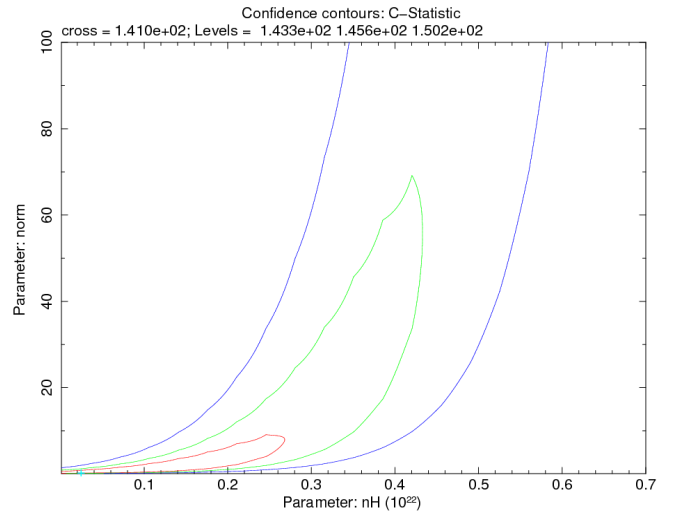}}
  \caption{Contour plot for the joint 2005-2007 fit of the column density $N_{\rm H}$ vs. blackbody normalization. The contours and symbols are the same as in Figure~\ref{fig:cont1}. The degeneracy between column density and blackbody normalization is evident. }\label{fig:cont2}. 
\end{figure}

We also try to leave the $N_{\rm\,H}$ value free in the fit, but the
strong degeneracy with the blackbody normalization $K$ (see
Figure~\ref{fig:cont2}) does allow to place only a 90\% confidence
upper limit of $N_{\rm H} < 4\times10^{21}\,$cm$^{-2}$. To better
constrain the surface of the emitting area we also try to fit a
neutron star atmosphere model (\emph{nsatmos}), where we fix the
neutron star mass, radius, distance, and column density at
$1.4\,$\msun, 10 km, 700 pc and $3\times10^{20}\rm\,cm^{-2}$,
respectively. We obtain a good fit for a hot cap with a temperature of
$1.1\times10^{6}$K and a 90\% confidence level upper limit on the
emitting area of less than $4\times10^{6}\rm\,m^2$. The results of
this model are shown in Table~\ref{tab:comb}

Following \citet{pav07} (see their Eq.~2), we can set an upper
limit on the putative cloud of absorbing material around the pulsar
from our constraints on the hydrogen column density. We assume that the
cloud has a spherical shape and that the volume is truncated at around
the location of the outermost planet ($\approx0.5\rm\,au$). This gives
an upper limit on the mass of the cloud of:
\begin{equation}\label{eq:masslost}
 M_{\rm cl}<1.5\times10^{-4}\left(\frac{\mathcal{L}}{0.5\rm\,au}\right)^2\rm\,M_{\oplus}
\end{equation}
at the 90\% confidence level, where $\mathcal{L}$ is the size of the cloud in units of au. For sizes of order 1 au and above, this similar to dusty debris disk masses around main-sequence stars~\citep{wya08}. Thus, if there is a dust cloud around B1257+12, it is comfortably in the mass regime of debris disks.

\section{Neutron star family}\label{sec:env}

%Type of neutron star systems
%For the sake of discussion, 
\subsection{Radio pulsars}\label{sec:pw}

Young PSRs are relatively young (0--1 Gyr) neutron stars formed after a
supernova explosion. Their spin period is usually in the 0.01--10 s
range and they possess a strong magnetic field of the order of
$10^{10}$--$10^{13}$ G.
The rotation of their strong dipolar magnetic field causes the
emission of low frequency radiation and a relativistic particle wind
with a loss of energy from the system with a consequent spin down. A
pulsar loses energy at a rate:
\begin{equation}
L_{\rm sd} = I\omega\dot\omega = 4\pi^2 I\frac{\dot{P}}{P^3}
  \end{equation}
where $L_{\rm sd}$ is the spin-down power, $I$ is the moment of inertia of the
neutron star, $\omega$ is the neutron
star angular frequency, $\dot\omega$ its spin down rate and $P$ and $\dot{P}$
are the spin period and the spin period derivative.

MSPs are old neutron stars that are formed in binaries and have been
spun up to millisecond spin periods by accretion processes (see
reviews \citealt{bha91, pat12r}). These objects can then either stop
accreting and become binary MSPs or ablate completely their companion
and become isolated MSPs.  Some formation scenarios for the pulsar
planets around \psrb\, discuss indeed their formation after the
millisecond pulsar ablated its companion leaving a post-ablation disk
of $\sim0.1\,$\msun~(\citealt{ste92}; see also \citealt{pod93} for a
review of different models proposed).  These pulsars are among the
most precise clocks in the Universe (e.g.,~\citealt{tay91}) and timing
their pulse arrival times allows the measurement of their spin and
orbital variations with exquisite precision. Thanks to this stability,
small perturbations that emerge in the timing analysis of MSPs can
highlight the presence of planets or other small bodies which would be
impossible to detect otherwise.

A large fraction of MSPs ($\sim80\%$)
are found in binaries with a companion star that can be another
neutron star, a main sequence star, a brown dwarf or a white dwarf.
Beside their spin period, MSPs differ from young PSRs also because
their dipolar magnetic field is of the order of $10^8-10^9$ G, whereas
the general physics of the pulsar wind and radiative emission is
believed to be similar.
Their weak magnetic field means also that they are spinning down at a
significantly slower pace when compared to the young pulsars, and
indeed most MSPs are expected to emit radio pulsations for at least a
few more billion years. Most of the energy lost is carried away by an
intense flux of low radio frequency radiation and a relativistic particle wind. A small
part of the energy budget ($\sim0.01-1\%$;~\citealt{pav07}) is instead
converted into high energy radiation consisting mostly of X-rays.

A large fraction of the spin-down power of pulsars is quickly converted into an
energetic flow of relativistic particles (pulsar wind), composed
mostly by electron-positron pairs.  The magnetic field of the pulsar
is instead dipolar (scaling with distance as $r^{-3}$) up to the
so-called light cylinder, which is the location where the magnetic
field lines of the pulsar have an angular velocity equal to the speed
of light.  After this point the field lines open and wind up creating
a toroidal field whose strength scales as $r^{-1}$.

The pulsar wind exerts a ram pressure $p_w$ that is a function of the total spin-down power of the pulsar (e.g.,~\citealt{har90}):
\begin{equation}\label{eq:pw}
 p_{w} = \xi\,L_{\rm sd}/(4\pi\,r^2\,c)
\end{equation}
where $c$ is the speed of light and $\xi$ is an efficiency factor that
accounts for the fraction of spin-down power that is effectively
transformed into relativistic wind power. Such parameter is different
than 1 because a fraction $0.01-0.8$ of the pulsar energy output
can be converted into high energy gamma ray photons, a discovery
recently made thanks to the \textit{Fermi} gamma-ray telescope (see
e.g.,~\citealt{abd13}). A small fraction of power is instead carried
away in form of X-rays and an even smaller fraction is transformed
into coherent radio waves detectable as radio
pulses from Earth, comprising only a tiny percentage of the energy
budget. Therefore realistic values for $\xi$ are in the 0.2-0.99
range. Such emission processes proceed until the pulsar crosses the
so-called ``death-line'', i.e., a region in the $P$-$\dot{P}$ diagram
where pulsars turn off since their spin-down power is insufficient
to sustain the pulsar wind and the production of electron-positron
pairs.

\subsection{Isolated neutron stars}

DINSs are also known as thermally emitting neutron
stars since their primary photon energy output comes from the thermal
emission from their surface. DINSs however, emit energy also in form
of neutrinos, with their cooling rate that can be described as:
\begin{equation}\label{eq:cooling}
  \frac{dU}{dt}=-L_\gamma-L_\nu
\end{equation}
where $U$ is the internal energy of the neutron star and $L_\gamma$ and $L_\nu$
are the photon and neutrino power. 
After a neutron star is formed, the neutrino energy loss is the dominant
mechanism, whereas photon cooling becomes dominant at a later stage.
To understand the typical timescale of this transition one needs to know
the exact interior composition of neutron stars since the neutrino
production strongly depends on the exact particle interactions that
occur in the neutron star core. For a number of plausible models,
the neutrino cooling dominates in the first 100,000 years (e.g., \citealt{pag06}).

A good approximation for $L_\gamma$ is given by blackbody emission:
\begin{equation}
L_\gamma = 4\pi\,R^2\sigma_{\rm\,SB}\,T_e^4. 
\end{equation}
where $T_e$ is the temperature of the neutron star envelope and $\sigma_{\rm SB}$ the Stefan-Boltzmann constant. 
If the neutron star is  in the photon cooling era, then $L_\nu\approx 0$ and the integration of Eq. (\ref{eq:cooling}) gives \citep{tsu79}:
\begin{equation}
t-t_0 = 2\times10^{3}\alpha^2\left(\frac{M_{\rm NS}}{M_{\odot}}\right)^{1/3}T_{e,7}^{-2}(f)\left[1-\left(\frac{T_{e,7}(f)}{T_{e,7}(i)}\right)^2\right]
\end{equation}
where the indices $f$ and $i$ indicate the final and initial
temperatures and $M_{\rm NS}$ is the neutron star mass (in the $\approx1.2-2.0$~\msun range. The parameter $\alpha\approx 0.1$--$1$ relates the
internal (core) temperature to the envelope temperature. The
$T\propto\,t^{-1/2}$ dependence of the temperature on time means that
the surface temperature halves as the age of the neutron star
quadruples. As an example, we start with a neutron star that has just
finished to cool mainly via neutrino emission, and we assume it has an
age of $0.1$ Myr. If the surface temperature of this object is of the
order of $10^5$ K (as observed in some isolated neutron stars of
similar age,~\citealt{kap08}) then its temperature would have
dropped to $\approx 6,000$ K when its age is ${\sim}1$ Gyr. The
thermal luminosity of the 1-Gyr old neutron star would then be
$L\approx10^{24}\rm\,\ergs$.

The minimum distance that a planet can have from the neutron star is the so-called tidal radius, which sets the region where an object would be disrupted by the tidal forces generated by the neutron star. This radius is:
\begin{equation}
  R_{\rm t} = \left(\frac{M_{\rm NS}}{M_{\rm\, p}}\right)^{1/3}R_{\rm\,p}.
\end{equation}
where $M_{\rm p}$ and $R_{\rm p}$ and the mass and radius of the
planet. The fraction of thermal radiation incident onto such planet
at the tidal radius distance is:
\begin{equation}
f = \frac{\pi\,R_{\rm p}^2}{4\pi\,R_{\rm t}^2} = \frac{1}{4}\left(\frac{R_{\rm\,p}}{R_{\rm t}}\right)^2 = \frac{1}{4}\left(\frac{M_{\rm\,p}}{M_{\rm NS}}\right)^{2/3}.
\end{equation}

For an Earth-like planet $f\approx 10^{-5}$ and thus the amount of
energy incident on the planet would be similar to that received by the
Earth from the Sun only if the neutron star temperature is slightly larger
than $10^6$ K. However, for a temperature of the neutron star of
$6,000$ K the amount of power received by the planet is equivalent to
what Earth would receive if it were at a distance of more than 100 au
from the Sun. Under these particular conditions planets will be frozen for most of their lifetime.

\subsection{Extra sources of heat and X-ray radiation}\label{sec:bondi}

Isolated neutron stars are directly exposed to the interstellar medium and it
is expected that all of them would accrete some of this material. Such
accretion process generates extra power due to the conversion of the
accreted gas rest mass into energy, with a typical efficiency of the
order of 10-20\%.This so-called Bondi-Hoyle accretion process should
be continuous and might be the main source of power for these type of
systems.~\citet{tor12} performed MHD calculations to estimate
the Bondi-Hoyle accretion luminosity of an isolated neutron star with
a magnetic field of $10^{12}\,$G (typical for PSRs) and found that $L_{\rm acc}=
\frac{G\,M_{\rm NS}\dot{M}}{R_{\rm\,NS}}\approx
6\times10^{29}\ergs$. If the neutron star has a magnetosphere, as
expected from most neutron stars, then this power might be mainly
emitted as X-ray photons pulsating at the rotational rate of the
neutron star.

In the case of pulsars, the X-ray radiation might also be formed as a
consequence of non-thermal processes originating in the pulsar
magnetosphere that heat up the polar caps (and which are unrelated to
accretion processes). To date few tens of MSPs
have been observed by the \textit{Chandra} and \textit{XMM-Newton}
observatories to emit 0.1-10 keV X-ray radiation with a power of the
order of $10^{29}-10^{31}\ergs$~\citep{pav07}. A few outliers
with a power of $10^{32}-10^{33}\ergs$ are also observed, but they
constitute a small minority of the sample.  Even if the X-ray
radiation represents a negligible fraction of the total power emitted
by pulsars, its effect on a planetary atmosphere might be severe given
that a substantial amount of energy can be deposited into the outer
layers of the atmosphere. Indeed the X-rays interact with the electron
K-shell of atoms, producing ions and energetic electrons
that drive the heating of the atmosphere~\citep{cec06}.
The fast electrons thus generated induce secondary ionizations that 
contribute substantially to the atmospheric chemistry and can provide
a substantial contribution to the atmospheric heating.

\begin{figure}
\rotatebox{-90}{\includegraphics[width=0.7\columnwidth]{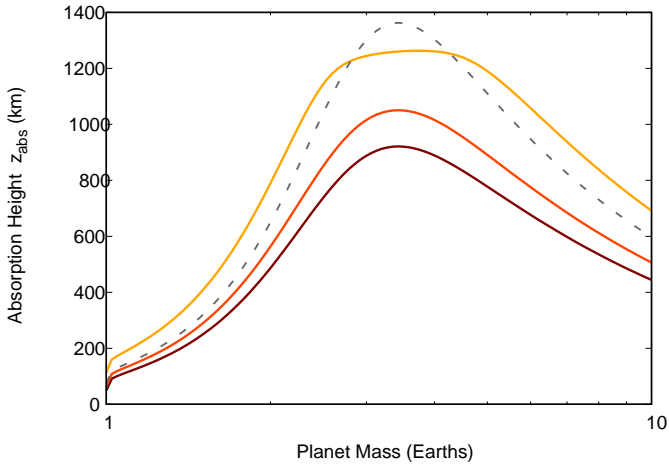}}
  \caption{Absorption height of high energy radiation for Earths and Super-Earths with total atmospheric thickness of $S=0.1~R_{\rm p}$. The solid curves show the penetration depth of radiation of different energy (hard X-rays/$\gamma$-rays $\beta_m=0.01\rightarrow$ dark red; hard X-rays $\beta_m=0.1\rightarrow$ orange; soft X-rays $\beta_m=100\rightarrow$ yellow). The dashed gray line identifies the threshold height where the number density reaches $10^{15}\rm\,cm^{-3}$. Below this threshold height, collisions play an important role in transferring heat to the lower atmosphere.}
  \label{fig:layer}
\end{figure}

\section{Effect of irradiation on neutron star planets}\label{sec:irr}

As we have discussed above, neutron stars can irradiate a planet via two main channels: relativistic winds (when the neutron star is active as a pulsar)
and/or via high energy X-ray radiation produced either during the pulsar phase
or via accretion from the interstellar medium for DINSs.

\subsection{Effect of pulsar winds}
For simplicity we begin by assuming that a planet around a young
pulsar possesses an isothermal atmosphere dominated by gas
pressure and no magnetosphere. 
The gas pressure is:
\begin{equation}\label{eq:pg}
  p_g=\rho\,k\,T/m
\end{equation}
where $T$ is the temperature of the atmosphere, $k$ is the Boltzmann
constant, $\rho$ is the density of the atmosphere which can be assumed
to have an exponential profile $\rho=\rho_0\,\rm\,exp(\it-z/h)$ and
$h=kT/m\,g$ is the scale-height (with $g$ being the gravitational
acceleration on the planet). The mean molecular mass is $m=\mu\,m_{\rm H}$
(where $\mu$ is the mean molecular weight and $m_{\rm H}$ is the hydrogen mass).
The pulsar wind and atmospheric pressures are equal at a height of about:
\begin{equation}\label{eq:z}
\hat{z} = h\,\rm\,ln\it\,\left(\frac{\rho_0\,k\,T\,4\,\pi\,r^2\,c}{\xi\,L_{\rm sd}\, m}\right)
\end{equation}

This is the location where the relativistic
particles that compose the wind will form a shock. The density after the shock
can be found by equating Eq.(\ref{eq:pw}) and (\ref{eq:pg}):
\begin{equation}
\rho=\frac{\xi\,L_{\rm sd} m}{4\pi\,r^2\,c\,k\,T}
\end{equation}

The temperature of the shocked gas can be calculated
from the perfect gas-law by knowing the pulsar-wind velocity $v_{\rm pw}$:
\begin{equation}
T_{\rm sh} = \psi\frac{m p}{k\,\rho}=\psi\frac{3}{16}\frac{m}{k}v_{\rm pw}^2.
  \end{equation}
where the parameter $\psi$ is an efficiency parameter which is equal
to one when the whole spin down power is carried by the relativistic
wind and the whole energy carried by the wind particles is transferred
into the atmospheric gas. In the remainder of the paper we assume for simplicity that $\psi\sim1$. Since the pulsar-wind is
ultra-relativistic, we can assume that $v_{\rm pw}\approx c$ and thus
the shocked gas will have an enormous temperature $T_{\rm
  sh}{\sim}10^{13}$ K.  The post-shock gas will be optically thin,
thus the emerging radiation will be characterized by a very small
Compton $y$-parameter and the main cooling mechanism for the shocked
gas will be thermal Bremsstrahlung. At these temperatures, the typical
photon energy is $\approx 1$ GeV (i.e., $\approx
1.6\times10^{-3}\rm\,erg$). Such high energy photons will penetrate deep into the atmosphere and deposit their energy in a layer whose depth depends on the photon energy (see Section~\ref{sec:he}).

Since the pulsar is injecting a fraction $f$ of its spin down energy into the
planet, then the mass loss $\dot{M}_{\rm pw}$ can reach its maximum allowed value if we assume that all the kinetic energy of the relativistic wind is transferred via collisions to the atmospheric particles (and thus we neglect the photon energy deposition discussed above):
\begin{eqnarray}\label{eq:mdotpw}
  &\dot{M}_{\rm pw}& = \psi\frac{f\,L_{\rm sd}}{U_{\rm b}} =\\\nonumber
&=&5\times10^{11}\psi\left(\frac{L_{\rm sd}}{10^{34}\rm\,erg\,s^{-1}}\right)\left(\frac{R_{\rm p}}{R_{\oplus}}\right)^{3}\left(\frac{D}{1\rm\,au}\right)^{-2}\left(\frac{M_{\rm p}}{5\,M_\oplus}\right)^{-1}\rm\,g\,s^{-1}
  \end{eqnarray}
where $U_{\rm b}$ is the specific gravitational binding energy,
$f=\frac{1}{4}\left(\frac{R_{\rm p}}{D}\right)^2$ depends on the
planet radius and distance ($D$) from the pulsar.
The total binding energy per unit mass on a planet is:
\begin{equation}
U_{\rm b} = -\frac{G M_{\rm p}}{R_{\rm p}}.
\end{equation}
By taking as a typical value the Earth's radius and $M_{\rm
  p}=5 M_{\oplus}$, then the binding energy is $U_{\rm
  b}\approx\,3\times10^{12}\rm\,erg\,g^{-1}$.  The high energy photons are much
more energetic than the typical mean molecular energy $U_{\rm
  b}m\approx\,5\times10^{-10}\rm\,erg$ and thus might potentially cause a
large mass outflow from the atmosphere.

\subsection{Effect of high energy radiation}\label{sec:he}

\citealt{smi04} and \citet{cec09} have shown that the X-ray
irradiation of the planetary atmospheres can produce a substantial
heating. Usually the high energy X-rays and gamma-rays can penetrate
much deeper in the atmosphere than the UV and soft X-rays. However,
for sufficiently thick atmospheres even the hardest X-ray/gamma
photons ($>10\rm\,keV$) might not reach the surface of the
planet. Those planets orbiting around isolated neutron stars (which
are not emitting pulsar winds, Section~\ref{sec:pw}) might therefore
undergo a similar evolution as the planets exposed to intense X-ray
radiation from young main sequence stars. On Earth, the flow of X-rays
is quickly blocked by the upper atmosphere (thermosphere) which is a
very optically thin layer of gas that becomes ionized when interacting
with X-rays and UV radiation. Such layer has a relatively large
temperature of hundreds up to thousands degrees but it is very
inefficient at conducting heat since it is very rarefied.  The
penetration depth of high energy radiation depends on three
quantities, the initial intensity $I_{\nu,0}$ of the radiation field,
the energy $E$ of the high energy photon and the composition of the
atmosphere. In general the attenuation of X-ray/gamma radiation is
characterized by the so-called linear attenuation coefficient $\beta$
and can be described as:
\begin{equation}\label{eq:I}
I_{\nu}(x) = I_{\nu\,0}e^{-\beta\,x}
\end{equation}
where $x$ is the distance traveled by the X-ray photon in the atmosphere. 
The linear attenuation coefficient is also dependent on the density of the
medium being crossed by the high energy radiation. Since the atmospheric density
scales exponentially, the linear attenuation coefficient can be written:

\begin{equation}\label{eq:mu}
\beta = \beta_m\,\rho_0\,\rm\,exp(\it\,-z/h)
\end{equation}
where $\beta_m$ is the mass attenuation coefficient and can be found in
tables\footnote{http://physics.nist.gov/xaamdi}. To find the height $z_{\rm abs}$ (measured from the planet surface) where the specific intensity 
will be reduced by a factor $e$, we define the total atmospheric thickness as $S$ such
that $x =S-z$ and combine Eq.(\ref{eq:I}) and (\ref{eq:mu}):
\begin{equation}
\rm\,ln\it\,\left(\frac{I_{\nu}}{I_{\nu,0}}\right)=-\beta_m\rho_0(S-z)e^{-z/h}
\end{equation}
The solution for this equation can be written as:
\begin{equation}
z_{\rm abs} = S- h\,W\left(\frac{e^{S/h}}{\rho_0\beta_m\,h}\right)
\end{equation}
where the function $W$ is the Lambert W function.  A heating of the
atmosphere by high energy radiation can penetrate in deeper
atmospheric layers than UV and/or soft X-ray radiation and thus
generate a hydrodynamic escape of the atmosphere, beside the thermal
Jeans escape due to direct collisions with atoms/molecules.

The typical density of the atmospheric layer where the X-ray photons
are absorbed will depend on a number of factors like the physical state and
chemical composition of the atmosphere. If we choose as an illustrative
example an Earth-like atmosphere, then the height $z_{\rm abs}$ will occur
at around 50-70 km from the surface, where we have used $\beta_m=0.1-0.01$
which is suitable for photons with energies larger than $\sim10$ keV. The number density can be calculated as:
\begin{equation}
n = \frac{\rho}{\mu\,m_{\rm H}}
\end{equation}
and gives a value of the order of $10^{17}\rm\,cm^{-3}$. At these
densities the metals in the atmosphere will undergo frequent
collisions and will be de-excited while being in turbulent or
thermally induced motion and transfer their energy to other
atoms/molecules.  An illustrative example obtained for Super-Earths of
different mass is shown in Figure~\ref{fig:layer}. The density of the
atmosphere at the surface of the planet is calculated by imposing
hydrostatic equilibrium. This implies that the atmospheres of larger
planets have a strong density gradient, whereas smaller planets have
more homogeneous atmospheres. Since Earth-like planets can retain only
a tiny fraction of atmospheric mass we have used a smoothing function
to connect the properties of Earth-like planets and Super-Earths.
The smoothing function is defined as:
\begin{equation}
f(M_p) = \rm\,tanh(\it\,M_p-M_o) + \epsilon
\end{equation}
where $M_0=1$~\mearth and $\epsilon=10^{-6}$ is the fraction of planetary mass
retained in the atmosphere of a 1~\mearth\, planet. Since there
might be also a strong compositional difference of the atmospheric
elements, we used a similar smoothing function for the molecular
weight (that we have chosen to go from $\mu=29$ for a 1~\mearth\, planet
down to 3 for a Super-Earth). Finally, the radius of the planet is
also increased gradually from 1~$R_{\oplus}$ (for an Earth like planet)
to 2~$R_{\oplus}$ for Super-Earths.

\section{Application to pulsar planetary systems}\label{sec:pla}
\begin{figure}
\includegraphics[width=1.0\columnwidth, angle=0]{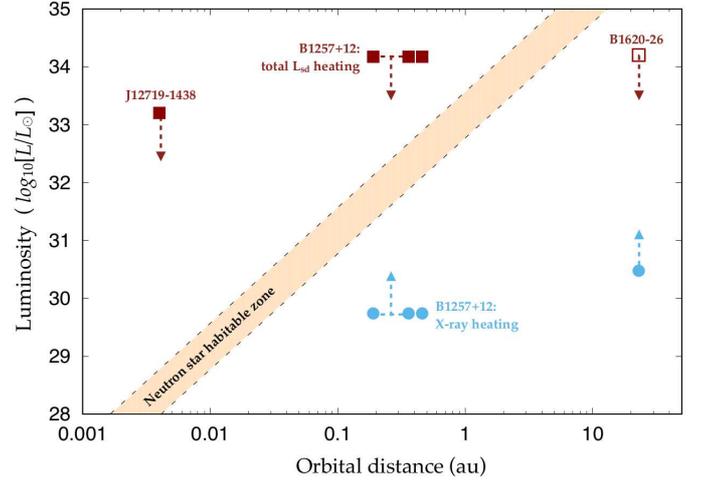}
\caption{Neutron star planets habitable zone. The diagram shows the
  luminosity produced by the neutron star (as pulsar wind or X-ray
  radiation), which is relevant to heat the planet atmosphere as a
  function of the planetary distance. The habitable zone is the orange
  area between the oblique dashed gray lines. Open symbols are upper limits.}
  \label{fig:hz}
\end{figure}
\begin{figure}[!]
\includegraphics[width=1.0\columnwidth]{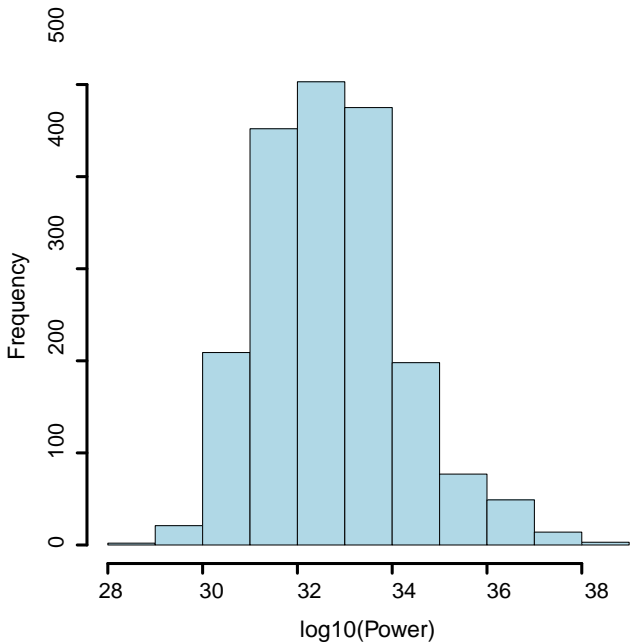}
  \caption{Histogram of the spin-down luminosity of radio pulsars.
The median value is $\approx4\times10^{32}\rm\,erg\,s^{-1}$.}
  \label{fig:dist}
\end{figure}

\subsection{Atmospheric heating from high energy radiation}\label{sec:hz}

% Definition for a normal star
There are several definitions in the literature about what constitutes
the boundaries of an habitable zone around main sequence stars. In
this work we follow the definition given by \citet{kas93}: the
habitable zone is the region around a star where a Earth-like planet
(with a $\rm\,CO_2/H_2O/N_2$ atmosphere) can have large amounts of
liquid water on its surface.

% Definition for a Neutron Star
According to \citet{sel07} a necessary (but not sufficient)
condition for habitability of a planet is that its equilibrium
temperature $T_{e\rm\,q}$ is below 270 K.  Such equilibrium
temperature can be defined as:
\begin{equation}\label{eq:teq}
T_{\rm\,eq} = \left[\frac{\Sigma(1-a)}{\phi\sigma}\right]^{1/4} 
\end{equation}
were $\Sigma$ is the incident stellar energy flux, $a$ is the Bond
albedo of the planet, $\sigma$ is the Stefan-Boltzmann constant and $\phi$ is a
geometric factor of order unity that accounts for the redistribution
of the heat on the surface.  This equilibrium temperature does not
relate necessarily to the surface temperature of the planet (see
\citealt{sel07} for a discussion). In this paper we follow
\citet{kal11} and set the inner and outer planet
habitable zone boundaries by requiring that the equilibrium
temperature lies in the range $175 K < T_{\rm eq} < 270 K$.

We can thus rewrite the above equilibrium temperature expression in
terms of the fraction $\eta$ of X-rays and gamma-rays which are able
to penetrate the atmosphere in a layer which is dense enough to allow
the thermalization of the released energy (rather than the thermal
escape of the gas).  If we rewrite Eq.~(\ref{eq:teq}) as a function of
the input neutron star luminosity ($L_{\rm NS}$) and distance of the
planet from the pulsar then we obtain:
\begin{equation}\label{eq:teq2}
T_{\rm\,eq} = \eta\left[\frac{L_{\rm NS}}{16\pi\sigma\,D^2}\right]^{1/4}. 
\end{equation}
The biggest uncertainty in this equation lies in the parameter $\eta$ which
depends on the energy of the photon and the physical parameters of the
atmosphere. 

\begin{figure*}[ht!]
  \begin{tabular}{cc}
    \includegraphics[width=0.5\textwidth]{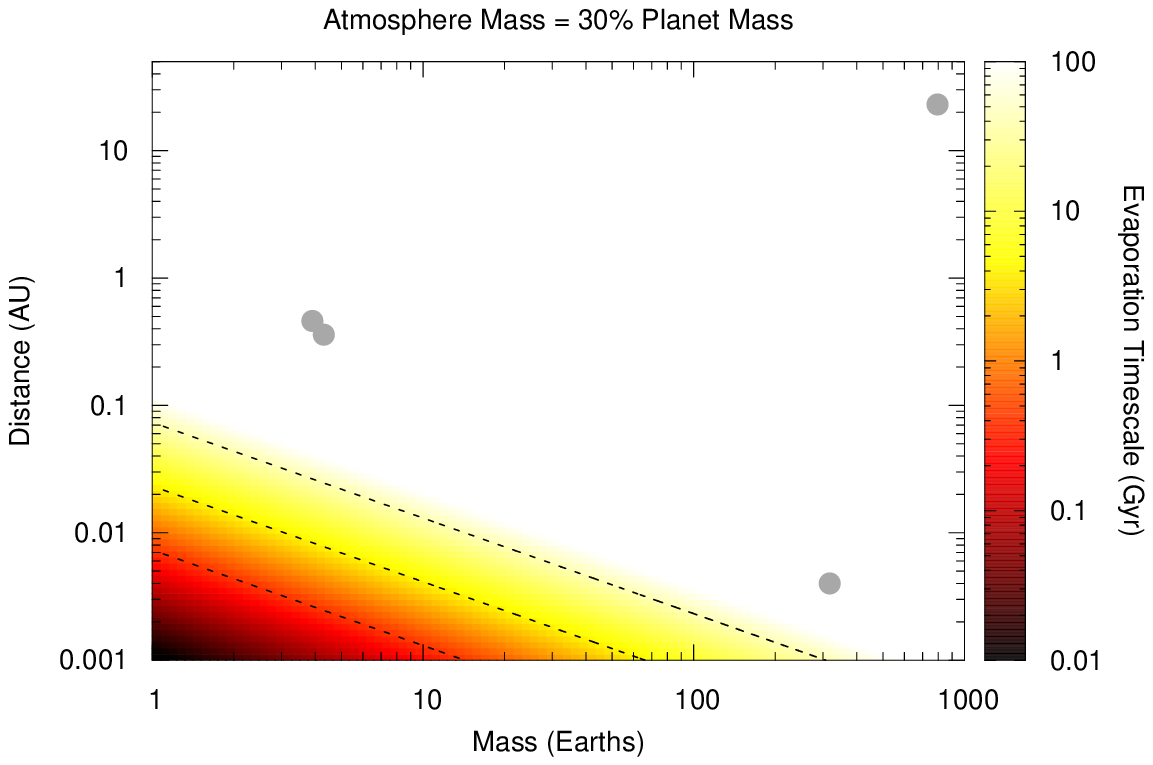}& \includegraphics[width=0.5\textwidth]{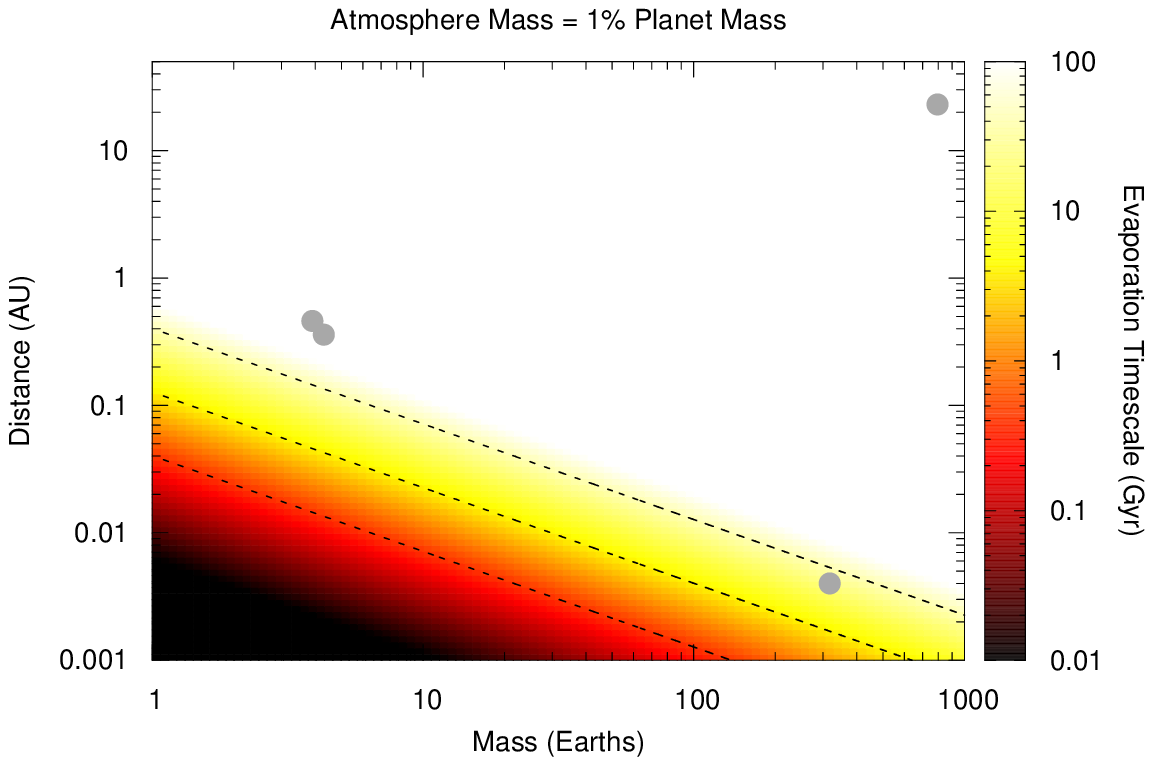}\\
    \includegraphics[width=0.5\textwidth]{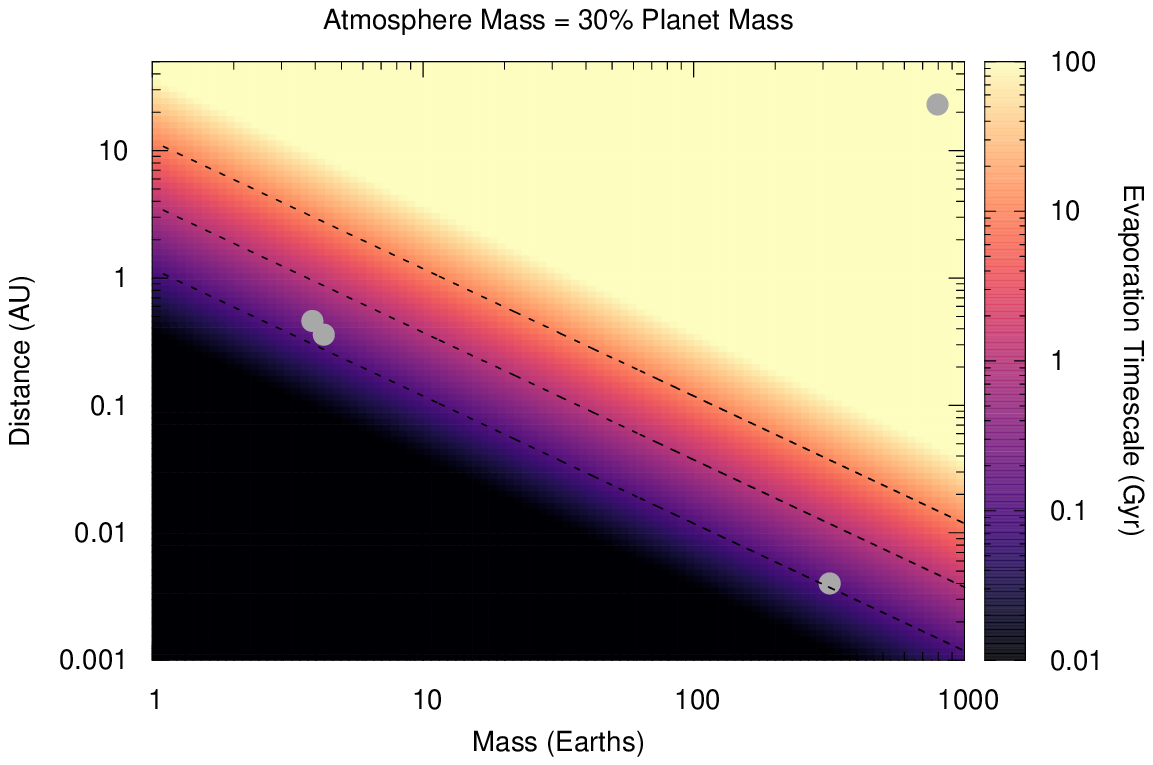}& \includegraphics[width=0.5\textwidth]{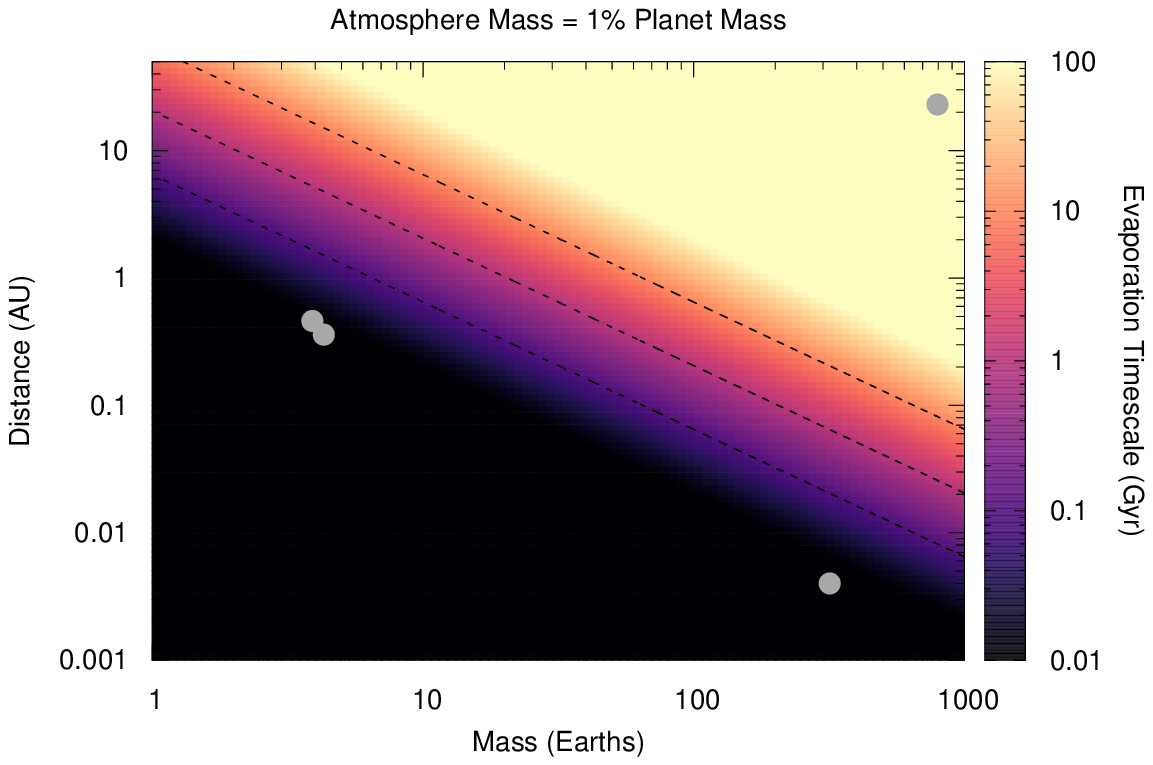}\\
    \includegraphics[width=0.5\textwidth]{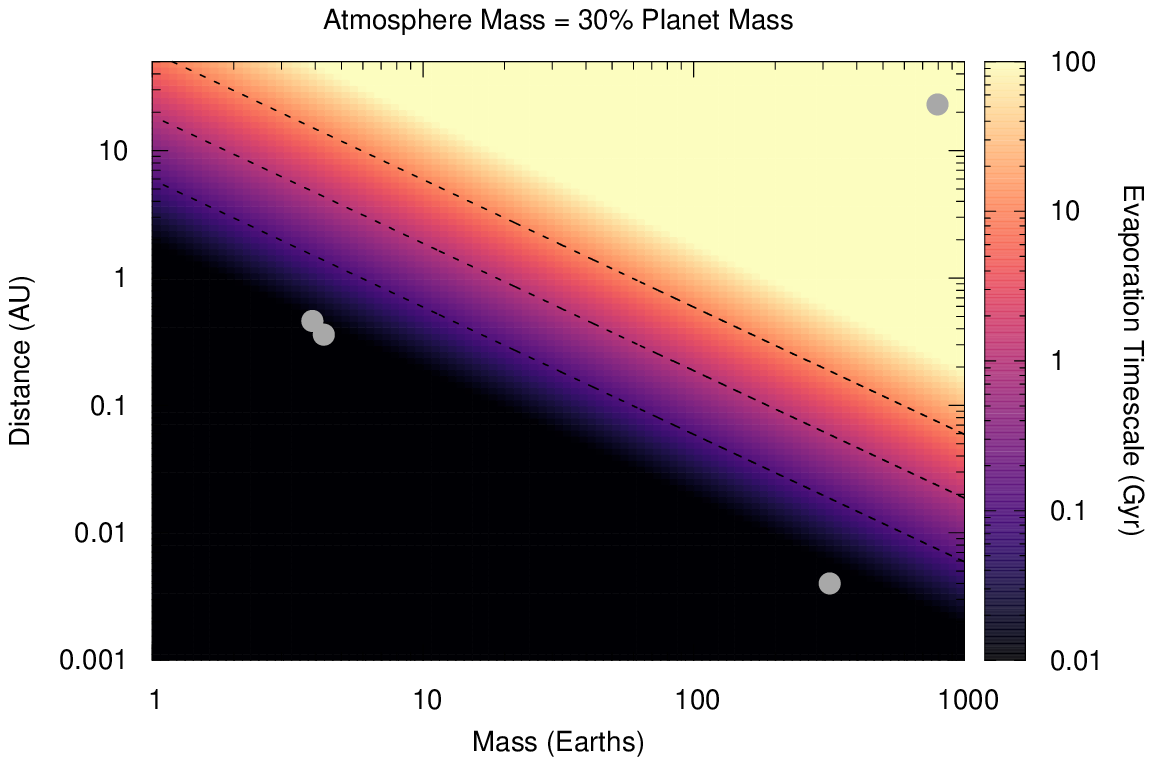}& \includegraphics[width=0.5\textwidth]{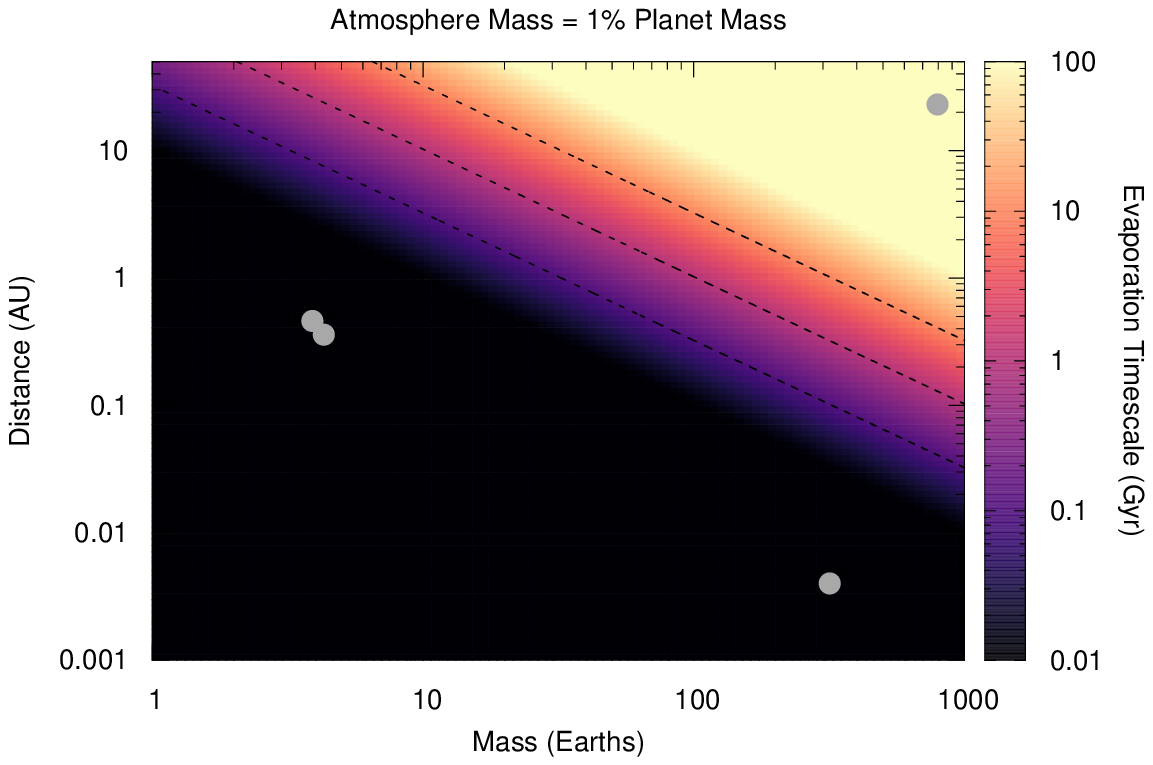}\\
  \end{tabular}
  \caption{Evaporation timescale of planets irradiated by a source of
    X-rays with $L_{\rm X}=10^{29}\ergs$ (top row), pulsar wind $L_{\rm
      sd}=4\times10^{32}\rm\,erg\,s$ (middle row) and $L_{\rm
      sd}=10^{34}\ergs$ (bottom row). All figures on the left show the case
    for a very thick atmosphere holding 30\% of the planet total
    mass. The right panels show the case for a thinner atmosphere of
    1\% of the total planet mass  (which is a
reasonable assumption for the atmospheric mass of
Super-Earths; see~\citealt{lop13,elk08a}). The gray circles represent the mass
    and distance of four of the five pulsar planets (with the
    exclusion of \psrb\, a, which is too small to hold any
    atmosphere. The dots represent (from left to right): \psrb\,b,
    \psrb\,c, PSR J1719–1438 b, PSR B1620-26 b. The dotted black lines
    represent isochrones of 0.1, 1 and 10 Gyr (from bottom to top,
    respectively). The radius of all planets has been fixed to $R_{\rm
      p}=2\,R_{\oplus}$. The evaporation timescales have been
    calculated with Eq.~\ref{eq:mdotX} (top figures)
    and~\ref{eq:mdotpw} (middle and bottom figures).}
  \label{fig:atm}
\end{figure*}

To have an equilibrium temperature between 175 and 270 K, a planet
needs to lie between the lower ($D_{\rm l}$) and upper ($D_{\rm u}$)
habitable zone boundaries. We start as an initial example with the
planets in the \psrb\, system where the total input neutron star
luminosity needs to be between $\eta\,L_{\rm X}\approx
\eta\,5.5\times10^{29}\rm\,erg\,s^{-1}$ (see \citealt{pav07,yan13} and
Section 3 of this work) and $L_{\rm
  sd}\approx1.5\times10^{34}\rm\,erg\,s^{-1}$. If the main source of
power comes from the X-ray luminosity alone, then even if we take
$\eta=1$ then $D_{\rm l} =0.02$ au and $D_{\rm u} = 0.06$ au and all
three pulsar planets in \psrb\ would be too cold.  However, if
sufficient gamma radiation is produced in the upper atmosphere due to
the presence of the pulsar wind shock, then the boundaries of the
habitable zone for planets b and c in \psrb\, shift outwards between 2
and 5 au (see Table~\ref{tab:psrb}). In Figure~\ref{fig:hz} we show
the neutron star habitable zone (calculated with Eq~\ref{eq:teq2})
defined as the location where a pulsar planet surface temperature is
between 175 K and 270 K.  The blue circles refer to the measured X-ray
luminosity and distance of the pulsar planets in PSR B1257+12 and PSR
B1620-26. The red squares (the open square is an upper limit) refer to
the total spin-down luminosity of the same pulsar planets with the
addition of the ``diamond planet'' pulsar PSR J1719–1438.  A plausible
irradiation luminosity for the planet lies between the red squares and
the blue circle if the X-ray luminosity is isotropic and the pulsar
wind (partially) hits the planet and the gamma-ray luminosity is
absorbed by the planet atmosphere. The planet around PSR B1620-26 is
too cold even in the most optimistic case, whereas the 175-270 K band
lies in between the values for the pulsar planets in PSR B1257+12. No
X-ray information is currently available for PSR J1719-1438 although
the X-ray luminosity required to fall in the temperate 175-270 K band
needs to be smaller than any other known X-ray pulsar.

\begin{table}[!ht]
\centering
\caption{Atmospheric Temperature Range for Pulsar Planets.}
\begin{tabular}{lccc}
\hline\hline
Planet & Mass & $T_{\rm min}$ & $T_{\rm max}$\\
   & ($M_{\oplus}$) & (K) & (K)\\
\hline
\psrb\, b & 0.02 & 70 & 899\\
\psrb\, c & 4.3 & 51 &  652\\
\psrb\, d & 3.9 & 45 &  577\\
PSR B1620--26 b & 795 & 10 &  83\\
PSR J1719--1438 b & 318 & NA & 3540 \\ 
\hline
\end{tabular}
\emph{\newline\,Note:} The planet \psrb b is probably not massive enough to retain any atmosphere.
\label{tab:psrb}
\end{table}

Furthermore most isolated neutron stars with Bondi-Hoyle accretion
have an X-ray luminosity larger than \psrb\, \citep{pfa01} which is on
the lower end of the X-ray luminosity spectrum rather than a typical
system.  For rocky planets similar to the Earth, such habitable zone
would exist for a very brief amount of time due to the atmospheric
evaporation whereas for Super-Earths with dense atmospheres this phase
could be potentially very long.
To evaluate the amount of matter lost by the planet's atmosphere as a
consequence of high energy irradiation we adapt the calculations of
\citet{lon81} and \citet{rud89}, who considered the evaporation rate
($\dot{M}_{\rm w}$) of a light stellar companion due to X-ray heating
from a pulsar.  Differently from \citet{rud89}, we do not use the
mass-radius relation for low mass stars, but we leave the radius and
mass dependencies of the planet explicitly in the equations.  We
assume that the velocity of the emerging particle wind due to X-ray
evaporation is equal to the escape velocity from the planet:
\begin{equation}
v_{\rm e} = \sqrt{\frac{GM_{\rm p}}{R_{\rm p}}} = 2.6\times\,10^{-4} \left(\frac{M}{R}\right)^{1/2}\rm\,cm\,s^{-1}
\end{equation}
For an Earth-like planet as the one under consideration here, the
total mass loss rate is therefore:
\begin{equation}\label{eq:mdotX}
\dot{M}_{\rm w} = 3\times 10^{-14}\chi\left(\frac{X_M}{10^{-3}}\right)\left(\frac{M_{\rm p}}{M_{\rm \oplus}}\right)^{-1/2}\left(\frac{R_{\rm p}}{R_{\rm \oplus}}\right)^{1/2}\frac{L_x}{\rm(erg/s)}\frac{R_{\rm p}^2}{4 D^2}\rm\,g\,s^{-1}
\end{equation}
where $X_M$ is the fractional metal abundance in the atmosphere, $\chi$ is the ratio
between the soft (0.2-1 keV) X-ray intensity to the total incident
X-ray intensity, relative to the same ratio for the X-ray binary
pulsar Her X-1 (which was used by \citealt{rud89} as a reference
source for the X-ray emission).  If we assume an atmosphere completely
dominated by heavy elements (as is the case for the Earth) and we use
$\chi=1$ then the expression above reduces to:
\begin{equation}
\dot{M}_{\rm w} \simeq 10^{7} \left(\frac{L_{\rm x}}{10^{30}\ergs}\right)\left(\frac{D}{1\rm\,au}\right)^{-2} \rm\,g\,s^{-1}
\end{equation} 
The evaporation timescale to lose an Earth-like atmosphere mass
($M_{\rm atm}$) would therefore be $\tau_{\rm evp} = M_{\rm atm}/\dot{M}_{\rm w}\sim 5\times10^{21}\rm\,g/\dot{M}_{\rm w}$.
    Typical values for $\tau_{\rm evp}$ lie in the range
    $10^6$--$10^7$ years for Earth-like planets at distances of
    $1$--$10$ au and up to trillion years for the most extreme
    Super-Earths with thick atmospheres (see Figure~\ref{fig:atm}).

\subsection{Habitable zone parameter range}

We can now ask the question about the physical parameters that a
planet needs to have in order to fall within the pulsar habitable zone
(as discussed in Section~\ref{sec:hz}).  We begin with a pulsar that
has a spin-down luminosity among the smallest possible values (see
Figure~\ref{fig:dist} for the distribution of spin-down luminosity for
all measured PSRs and MSPs), i.e.  $L_{\rm sd}\approx
10^{29}\rm\,erg\,s^{-1}$ and we consider the effect of the wind on its
atmosphere. % (see Figure~\ref{fig:ppdot}).  If we assume that the
planet is similar to the Earth (in mass, atmospheric composition,
density and distance) then the shock will form at an height $\hat{z}$
of approximately 200 km (see Eq.~\ref{eq:z}) which would correspond to
the thermosphere of the planet.  We now require that the planet
equilibrium temperature lies within the range 175--275 K and we use
Eq.~(\ref{eq:teq}) to find a range of allowed distances: 0.02--0.06
au.  Using Eq.~(\ref{eq:mdotpw}) for a typical Earth mass/radius
planet, gives a mass loss rate of the order of $10^{12}\rm\,g\,s^{-1}$
which means that an Earth-like atmosphere would be potentially
consumed in a time as short as few hundred years (for spin down luminosities
of $10^{34}\rm\,erg\,s^{-1}$ . The most optimistic
case that will give the longest possible survival time of the
atmosphere is the case of a Super-Earth which, in the most extreme
cases, can have an atmosphere of about 30\% the planet's mass
\citep{elk08a}. The binding energy $U_{\rm b}$ for such planet would
be similar to that of an Earth-like
planet and the corresponding mass loss would be $\dot{M}_{\rm
  pw}\approx 10^{12}\rm\,g\,s^{-1}$. The atmospheric mass of such
planets can reach up to ${\sim}10^{28}\rm\,g$, which would give an
evaporation timescale of the order of 0.1-1 Gyr (assuming that all
pulsar wind power is used to eject the atmospheric gas).

It is important also to stress that the pulsar wind is not a process that will continue
indefinitely.  Indeed once the pulsar reaches a sufficiently slow spin
period it will cross the so-called ``death-line'' in the $P-\dot{P}$
diagram, meaning that the pulsar wind will turn off. For young
pulsars, this occurs on a timescale of the order of million years,
whereas for MSPs it is of the order of billion years. However, this
would also turn off the energy source of the planet and therefore the
temperature will drop dramatically thus removing any possibility to
define an habitable zone, unless a Bondi-Hoyle accretion process
generates a sufficiently large amount of X-ray radiation as discussed
in Section~\ref{sec:bondi} or other effects like radiogenic heating (Section~\ref{sec:comp}) or tidal heating play a dominant role.

\subsection{Presence of a magnetosphere}\label{sec:planetmagnet}

If a magnetosphere is present around pulsar planets, then its shielding effect can be fundamental to deflect and/or confine the incoming relativistic pulsar wind particles. If we assume that the planetary magnetic field $B_{\rm p}(r)$ is dipolar, then when the magnetic pressure balances the pulsar wind ram pressure, a shock will form at a distance $r_{\rm s}$ from the planet:
\begin{equation}
  \frac{B_{\rm p}^2(r_s)}{8\pi}= \xi\,L_{\rm sd}/(4\pi\,r_s^2\,c).
\end{equation}
If $B_{\rm p}$ is the value of the magnetic field on the planet's surface, then $r_s$ can be found imposing the condition that it is larger than the radius of the planet $R_{\rm p}$. 
In Table~\ref{tab:B} we calculate the minimum surface magnetic field $B_{\rm p, min}$ required for each of the known pulsar planets to withstand the ram pressure of the pulsar wind.

\begin{table}[!ht]
\centering
\caption{Minimum planetary surface magnetic field required to prevent atmosphere loss by pulsar wind stripping.}
\begin{tabular}{lcccc}
\hline\hline
Planet System & Mass & D & $L_{\rm sd}$ & $B_{\rm p,min}$\\
   & ($M_{\oplus}$) & (au) & $10^{34}\ergs$ & G\\
\hline
\psrb\, b & 0.020 & 0.19 & 1.5 & 0.35\\
\psrb\, c & 4.3 & 0.36 & 1.5& 0.19\\
\psrb\, d & 3.9 & 0.46 & 1.5& 0.15\\
PSR B1620--26 b & 795 & 23 & <1.9 & <0.003\\
PSR J1719--1438 b & 318 & 0.004 & 0.16 & 5.6\\
\hline
Earth	& 1	& 1	& --	& 0.22-0.67\\
\hline
\end{tabular}
\emph{Notes:} Calculations assume $\xi=1$. Earth is given for reference.
\label{tab:B}
\end{table}

\subsection{Are pulsar winds really hitting the planets?}

In the preceding discussion we have assumed that the pulsar wind is
emitted isotropically. The mechanism to produce a radio pulsar beam
and particle wind is still not completely understood, but the current
physical picture requires the production of electron-positron pairs.
When pair creation happens, the negative current flows along the poles
whereas the positive charge/current flows in a sheet on the pulsar
equator~\citep{che14}.  In the open filed zone there is a jump in the
toroidal field component $B_{\phi}$ that requires charge density to be
negative~\citep{lyu90}. There is a matter flow of ions in the
equatorial plane with energy density which is about twice the magnetic
field energy density.  In the aligned rotator geometry (i.e., when
spin and magnetic axes are nearly aligned), the current sheet of ions
in the equator extends indefinitely if nothing stops this particle
flux. If the misalignment is instead substantial, the flow of
particles oscillates~\citep{phi15,tch16}.
It is thus possible to speculate that the wind could completely miss
the planets orbiting around the pulsar because of a geometric
misalignment of the orbital plane with respect to the ion current
sheet or negative charge flow. In this case the planet will be only
irradiated by the X-ray emission of the neutron star.

For \psrb\, we can calculate with Eq.~\ref{eq:mdotX} what would be the
total mass lost by the planets in this case. By using the current
characteristic age of the pulsar ($\approx850$ Myr) we expect that the
two outermost planets should have lost a total of
$\approx5\times10^{-4}M_{\oplus}$.  We caution that given the many
uncertainties on the quantities involved, it is not possible to
conclude that this is the correct scenario for \psrb\,. However, we
note that the value calculated above is close to the
estimate made in Eq.~\ref{eq:masslost} based on X-ray observations.

\section{Formation and composition of neutron star planets}\label{sec:comp}

The low-mass planetary system around PSR~B1257+12 is unique in its
architecture among the ${\sim}400$ millisecond pulsars\footnote{See
  https://apatruno.wordpress.com/about/millisecond-pulsar-catalogue/
  for an up-to-date catalog of MSPs}. This has been interpreted as
evidence for a very low probability formation channel. However,
low-mass planetary systems at larger orbital separations are still
unconstrained around MSPs and other neutron stars. Therefore, the
PSR~B1257+12 planets may represent a rare case of either forming at,
or migrating to, sub-au distances, while planet formation around
neutron stars could still be a general phenomenon.

Neutron star planets may originate from one of three epochs: a
protoplanetary disk during the star formation process (1st
generation); a post-supernova fallback disk (2nd gen.); or an
accretion disk formed from material stripped by the neutron star from
a binary companion (3rd gen.). Any 1st generation planets, if they
exist, are likely destroyed or have their orbits disrupted during the
supernova. Millisecond pulsar planets cannot easily be explained by
2nd generation formation, as spin-up of the pulsar by a companion star
which migrated to within $1\,$au would disrupt their orbits.
This is justified by the fact that, since \psrb\, contains a fully recycled pulsar, its companion star
must have been a low-mass star which has undergone Roche lobe overflow during the main sequence phase. This requires that the orbital separation had been less
than about $\sim1$au. 
Thus, 3rd
generation formation, from the remnants of the companion star, is
preferred \citep{pod93}.

The gas-to-dust ratio, $\Delta_{\rm g/d}$, i.e. the mass ratio of
gas-phase volatiles (H, He, N, etc) to solids (Fe, Mg, Si, Al etc.) is
a major controlling parameter of planet formation. Models of the
formation of the PSR~B1257+12 planets have found that protosolar
nebula like conditions with $\Delta_{\rm g/d}=100$ are preferred over
a supernova fallback disk with $\Delta_{\rm g/d}\approx4$
\citep[e.g.][]{CurrieHansen2007}. Indeed, the architecture of the
PSR~B1257+12 system resembles the inner solar system, and it may have
formed from a radially confined planetesimal belt similar to that
proposed for the solar system \citep{Hansenetal2009}.

Based on main sequence and supernova
nucleosynthesis models for progenitor masses ${\leq}25\,$M$_{\rm
  \odot}$, the gas-to-dust ratio of post-supernova material is
${\gtrsim}20$ \citep{nom06}. Assuming a sufficiently massive (M$_{\rm
  dust}{\gtrsim}10^{-4}\,$M$_{\rm \odot}$) fallback disk remains
around the newly-formed neutron star, the high gas-to-dust ratio
would speed up dust growth and favour the occurrence of the streaming
instability ($\Delta_{\rm g/d}\gtrsim1$), potentially leading to a rapid build-up
of planetesimals \citep{joh07}. Such post-supernova material would also have a peculiar composition. The mass ratio of O to Si, Fe and Mg in the supernova ejecta increases from roughly solar ($\approx$unity) for a $13\,$M$_{\rm \odot}$ progenitor to $\sim3$ for $25\,$M$_{\rm \odot}$ \citep{nom06}. The fraction of O left over from rock formation and available for H$_{\rm 2}$O formation is different in these cases, so 2nd generation neutron star planets could potentially be very water-rich.

A major difference between neutron stars and main sequence ones is the
shape of their radiation field. Radio waves would not participate
significantly in disk heating or ionisation, and the ultraviolet and
visible luminosity is negligible for evolved systems although it may
play a role when the disk is accreting onto the neutron star
\citep[e.g.][]{mar16}.  The high flux of high-energy photons
and relativistic wind particles may lead to a high ionization fraction
which could mean there is no dead zone in the
disk. \citet{mar16} suggest that this would mean a low
formation efficiency of rocky planets, as dead zones are thought to
aid growth \citep{Johansenetal2014}. As discussed earlier, the shape
of the relativistic particle wind is currently not known, and its role
in the disk ionisation is thus not well constrained, so we leave these
issues open.

\begin{table}[!ht]
\centering
\caption{Number abundance of and energy production due to \fortyK.}
\begin{tabular}{ c c c }
\hline\hline
Type					&	$X$(\fortyK)	&	($10^{-8}\rm\,erg\,s^{-1}\,g^{-1}$)	\\
					&	$\times10^{-7}$	&	at $4.6\,$Gyr				\\
\hline
Earth				&	$0.95$		&	$2.9$					\\
SNII $15\,$M$_{\rm \odot}$	&	$2.48$		&	$7.6$					\\
SNII $20\,$M$_{\rm \odot}$	&	$326.5$		&	$996.7$					\\
\hline
\end{tabular}
\label{tab:fortyK}
\newline
\emph{Notes: }Number abundance is calculated as \fortyK\ atoms relative to total refractory atoms, approximated by including C, O, Mg, Fe, Si, Al, and S. Supernova abundances are from \citet{rau02}.
\end{table}

\subsection{Evolution of internal temperature: radiogenic heating}

The internal energy flux of Earth is $(47\pm2)\times10^{19}\rm\,erg\,s^{-1}$ or
$7.87\times10^{-8}\rm\,erg\,s^{-1}\,g^{-1}$ \citep{DaviesDavies2010}. This
energy comes from radioactive isotopes, currently totalling about
$5\times10^{-8}\rm\,erg\,s^{-1}\,g^{-1}$, of which
$2.9\times10^{-8}\rm\,erg\,s^{-1}\,g^{-1}$ is from \fortyK\ and the rest
largely \eightU\ and \twoTh, and from the heat of formation. For
Earth, the total internal energy production is only $0.027\,$\% of
solar insolation and does not directly control the surface
temperature. It is, however, important for habitability because it
drives plate tectonics and the geodynamo which gives rise to the
planetary magnetic field. This may offer protection against atmospheric ablation by the pulsar wind (Section~\ref{sec:planetmagnet}).

Second-generation planets around a neutron star would be composed of
the ejecta of the progenitor supernova, which can produce a wide range
of radioisotope abundances. Notably, \eightU\ and \twoTh\ may be less 
relevant in SNII ejecta than in average galactic material, because they are 
thought to be produced in neutron star mergers rather than Type~II supernovae,
although this is still being debated \citep{Eichleretal1989, Freiburghausetal1999, 
Cowanetal2005, Tanviretal2013}. The production of the long-lived planetary 
energy source \fortyK\ in SNII is well established \citep[e.g.][]{rau02,
kob06}. Table~\ref{tab:fortyK} shows the energy
production per gram of planetary core (refractory) mass for Earth
and for planets composed of ejecta from SNII with different progenitor
masses.

\section{Conclusions}

We have shown that the harsh environment around neutron stars can
still accommodate planets with warm atmospheres, provided they are
Super-Earths.  We addressed the effect of radiation produced by
neutron stars on planetary atmospheres and defined a habitable zone
for the particular case of low UV/optical radiation, moderate X-ray,
and high gamma-ray and relativistic particle flux.  We find that, if
part of the pulsar power is injected in the atmosphere, all of the
three PSR B1257+12 planets may lie in the habitable zone. In
particular, the two Super-Earths may have retained their atmosphere
for at least a hundred million years provided they contain a large
atmospheric fraction of the total planet mass, with the atmosphere
possibly still being present to these days.  We also find that if a
moderately strong planetary magnetosphere is present, the atmospheres
can survive the strong pulsar winds and reach survival timescales of
several billion years. The same argument applies to possible pulsar
planets around more powerful objects than \psrb\,.  Alternatively, a
similar result can be achieved if a non-isotropic pulsar wind is
present in the system.  Furthermore, planets which lie within a band
of $\approx0.01-1$ au can be in their habitable zone provided that the
neutron stars transfer their energy to the planetary atmosphere via
X-rays alone (e.g., through Bondi-Hoyle accretion).
We have also briefly discussed the formation of neutron star planets,
their potential high water content, and we have highlighted the fact
that radiogenic heating could play an important role in these systems.

\section*{Acknowledgments}
We would like to thanks O. Shorttle, Y. Cavecchi, M. Kenworthy, J. Hessels, R. Wijnands, S. Portegies Zwart, B. Czerny and T. Stolker for useful discussions. A.P. acknowledges support from a NWO Vidi Fellowship.

\bibliographystyle{aa}
\bibliography{ms}

\begin{thebibliography}{60}
\expandafter\ifx\csname natexlab\endcsname\relax\def\natexlab#1{#1}\fi

\bibitem[{{Abdo} {et~al.}(2013){Abdo}, {Ajello}, {Allafort}, {Baldini},
  {Ballet}, {Barbiellini}, {Baring}, {Bastieri}, {Belfiore}, {Bellazzini}, \&
  et~al.}]{abd13}
{Abdo}, A.~A., {Ajello}, M., {Allafort}, A., {et~al.} 2013, \apjs, 208, 17

\bibitem[{{Agol}(2011)}]{ago11}
{Agol}, E. 2011, \apjl, 731, L31

\bibitem[{{Bailes} {et~al.}(2011){Bailes}, {Bates}, {Bhalerao}, {Bhat},
  {Burgay}, {Burke-Spolaor}, {D'Amico}, {Johnston}, {Keith}, {Kramer},
  {Kulkarni}, {Levin}, {Lyne}, {Milia}, {Possenti}, {Spitler}, {Stappers}, \&
  {van Straten}}]{bai11}
{Bailes}, M., {Bates}, S.~D., {Bhalerao}, V., {et~al.} 2011, Science, 333, 1717

\bibitem[{{Batalha} {et~al.}(2013){Batalha}, {Rowe}, {Bryson}, {Barclay},
  {Burke}, {Caldwell}, {Christiansen}, {Mullally}, {Thompson}, {Brown},
  {Dupree}, {Fabrycky}, {Ford}, {Fortney}, {Gilliland}, {Isaacson}, {Latham},
  {Marcy}, {Quinn}, {Ragozzine}, {Shporer}, {Borucki}, {Ciardi}, {Gautier},
  {Haas}, {Jenkins}, {Koch}, {Lissauer}, {Rapin}, {Basri}, {Boss}, {Buchhave},
  {Carter}, {Charbonneau}, {Christensen-Dalsgaard}, {Clarke}, {Cochran},
  {Demory}, {Desert}, {Devore}, {Doyle}, {Esquerdo}, {Everett}, {Fressin},
  {Geary}, {Girouard}, {Gould}, {Hall}, {Holman}, {Howard}, {Howell},
  {Ibrahim}, {Kinemuchi}, {Kjeldsen}, {Klaus}, {Li}, {Lucas}, {Meibom},
  {Morris}, {Pr{\v s}a}, {Quintana}, {Sanderfer}, {Sasselov}, {Seader},
  {Smith}, {Steffen}, {Still}, {Stumpe}, {Tarter}, {Tenenbaum}, {Torres},
  {Twicken}, {Uddin}, {Van Cleve}, {Walkowicz}, \& {Welsh}}]{bat13}
{Batalha}, N.~M., {Rowe}, J.~F., {Bryson}, S.~T., {et~al.} 2013, \apjs, 204, 24

\bibitem[{{Bhattacharya} \& {van den Heuvel}(1991)}]{bha91}
{Bhattacharya}, D. \& {van den Heuvel}, E.~P.~J. 1991, PhysRep, 203, 1

\bibitem[{{Bisnovatyi-Kogan}(1992)}]{kog92}
{Bisnovatyi-Kogan}, G.~S. 1992, in IAU Symposium, Vol. 149, The Stellar
  Populations of Galaxies, ed. B.~{Barbuy} \& A.~{Renzini}, 379

\bibitem[{{Borucki} {et~al.}(2010){Borucki}, {Koch}, {Basri}, {Batalha},
  {Brown}, {Caldwell}, {Caldwell}, {Christensen-Dalsgaard}, {Cochran},
  {DeVore}, {Dunham}, {Dupree}, {Gautier}, {Geary}, {Gilliland}, {Gould},
  {Howell}, {Jenkins}, {Kondo}, {Latham}, {Marcy}, {Meibom}, {Kjeldsen},
  {Lissauer}, {Monet}, {Morrison}, {Sasselov}, {Tarter}, {Boss}, {Brownlee},
  {Owen}, {Buzasi}, {Charbonneau}, {Doyle}, {Fortney}, {Ford}, {Holman},
  {Seager}, {Steffen}, {Welsh}, {Rowe}, {Anderson}, {Buchhave}, {Ciardi},
  {Walkowicz}, {Sherry}, {Horch}, {Isaacson}, {Everett}, {Fischer}, {Torres},
  {Johnson}, {Endl}, {MacQueen}, {Bryson}, {Dotson}, {Haas}, {Kolodziejczak},
  {Van Cleve}, {Chandrasekaran}, {Twicken}, {Quintana}, {Clarke}, {Allen},
  {Li}, {Wu}, {Tenenbaum}, {Verner}, {Bruhweiler}, {Barnes}, \& {Prsa}}]{bor10}
{Borucki}, W.~J., {Koch}, D., {Basri}, G., {et~al.} 2010, Science, 327, 977

\bibitem[{{Brook} {et~al.}(2014){Brook}, {Karastergiou}, {Buchner}, {Roberts},
  {Keith}, {Johnston}, \& {Shannon}}]{bro14}
{Brook}, P.~R., {Karastergiou}, A., {Buchner}, S., {et~al.} 2014, \apjl, 780,
  L31

\bibitem[{{Burke} {et~al.}(2014){Burke}, {Bryson}, {Mullally}, {Rowe},
  {Christiansen}, {Thompson}, {Coughlin}, {Haas}, {Batalha}, {Caldwell},
  {Jenkins}, {Still}, {Barclay}, {Borucki}, {Chaplin}, {Ciardi}, {Clarke},
  {Cochran}, {Demory}, {Esquerdo}, {Gautier}, {Gilliland}, {Girouard}, {Havel},
  {Henze}, {Howell}, {Huber}, {Latham}, {Li}, {Morehead}, {Morton}, {Pepper},
  {Quintana}, {Ragozzine}, {Seader}, {Shah}, {Shporer}, {Tenenbaum}, {Twicken},
  \& {Wolfgang}}]{bur14}
{Burke}, C.~J., {Bryson}, S.~T., {Mullally}, F., {et~al.} 2014, \apjs, 210, 19

\bibitem[{{Campana} {et~al.}(2011){Campana}, {Lodato}, {D'Avanzo}, {Panagia},
  {Rossi}, {Della Valle}, {Tagliaferri}, {Antonelli}, {Covino}, {Ghirlanda},
  {Ghisellini}, {Melandri}, {Pian}, {Salvaterra}, {Cusumano}, {D'Elia},
  {Fugazza}, {Palazzi}, {Sbarufatti}, \& {Vergani}}]{cam11}
{Campana}, S., {Lodato}, G., {D'Avanzo}, P., {et~al.} 2011, \nat, 480, 69

\bibitem[{{Cash}(1979)}]{cas79}
{Cash}, W. 1979, \apj, 228, 939

\bibitem[{{Cecchi-Pestellini} {et~al.}(2006){Cecchi-Pestellini}, {Ciaravella},
  \& {Micela}}]{cec06}
{Cecchi-Pestellini}, C., {Ciaravella}, A., \& {Micela}, G. 2006, \aap, 458, L13

\bibitem[{{Cecchi-Pestellini} {et~al.}(2009){Cecchi-Pestellini}, {Ciaravella},
  {Micela}, \& {Penz}}]{cec09}
{Cecchi-Pestellini}, C., {Ciaravella}, A., {Micela}, G., \& {Penz}, T. 2009,
  \aap, 496, 863

\bibitem[{{Chen} \& {Beloborodov}(2014)}]{che14}
{Chen}, A.~Y. \& {Beloborodov}, A.~M. 2014, \apjl, 795, L22

\bibitem[{{Cowan} {et~al.}(2005){Cowan}, {Sneden}, {Beers}, {Lawler},
  {Simmerer}, {Truran}, {Primas}, {Collier}, \& {Burles}}]{Cowanetal2005}
{Cowan}, J.~J., {Sneden}, C., {Beers}, T.~C., {et~al.} 2005, \apj, 627, 238

\bibitem[{{Currie} \& {Hansen}(2007)}]{CurrieHansen2007}
{Currie}, T. \& {Hansen}, B. 2007, \apj, 666, 1232

\bibitem[{Davies \& Davies(2010)}]{DaviesDavies2010}
Davies, J.~H. \& Davies, D.~R. 2010, Solid Earth, 1, 5

\bibitem[{{Eichler} {et~al.}(1989){Eichler}, {Livio}, {Piran}, \&
  {Schramm}}]{Eichleretal1989}
{Eichler}, D., {Livio}, M., {Piran}, T., \& {Schramm}, D.~N. 1989, \nat, 340,
  126

\bibitem[{{Elkins-Tanton} \& {Seager}(2008)}]{elk08a}
{Elkins-Tanton}, L.~T. \& {Seager}, S. 2008, \apj, 685, 1237

\bibitem[{{Freiburghaus} {et~al.}(1999){Freiburghaus}, {Rosswog}, \&
  {Thielemann}}]{Freiburghausetal1999}
{Freiburghaus}, C., {Rosswog}, S., \& {Thielemann}, F.-K. 1999, \apjl, 525,
  L121

\bibitem[{{Fryer} \& {Heger}(2000)}]{FryerHeger2000}
{Fryer}, C.~L. \& {Heger}, A. 2000, \apj, 541, 1033

\bibitem[{{Geng} \& {Huang}(2015)}]{gen15}
{Geng}, J.~J. \& {Huang}, Y.~F. 2015, \apj, 809, 24

\bibitem[{{Hansen} {et~al.}(2009){Hansen}, {Shih}, \&
  {Currie}}]{Hansenetal2009}
{Hansen}, B.~M.~S., {Shih}, H.-Y., \& {Currie}, T. 2009, \apj, 691, 382

\bibitem[{{Harding} \& {Gaisser}(1990)}]{har90}
{Harding}, A.~K. \& {Gaisser}, T.~K. 1990, \apj, 358, 561

\bibitem[{{Johansen} {et~al.}(2014){Johansen}, {Blum}, {Tanaka}, {Ormel},
  {Bizzarro}, \& {Rickman}}]{Johansenetal2014}
{Johansen}, A., {Blum}, J., {Tanaka}, H., {et~al.} 2014, Protostars and Planets
  VI, 547

\bibitem[{{Johansen} {et~al.}(2007){Johansen}, {Oishi}, {Mac Low}, {Klahr},
  {Henning}, \& {Youdin}}]{joh07}
{Johansen}, A., {Oishi}, J.~S., {Mac Low}, M.-M., {et~al.} 2007, \nat, 448,
  1022

\bibitem[{{Kaltenegger} \& {Sasselov}(2011)}]{kal11}
{Kaltenegger}, L. \& {Sasselov}, D. 2011, \apjl, 736, L25

\bibitem[{{Kaplan}(2008)}]{kap08}
{Kaplan}, D.~L. 2008, in American Institute of Physics Conference Series, Vol.
  983, 40 Years of Pulsars: Millisecond Pulsars, Magnetars and More, ed.
  C.~{Bassa}, Z.~{Wang}, A.~{Cumming}, \& V.~M. {Kaspi}, 331--339

\bibitem[{{Kaplan} \& {van Kerkwijk}(2009)}]{kap09}
{Kaplan}, D.~L. \& {van Kerkwijk}, M.~H. 2009, \apj, 705, 798

\bibitem[{{Kasting} {et~al.}(1993){Kasting}, {Whitmire}, \& {Reynolds}}]{kas93}
{Kasting}, J.~F., {Whitmire}, D.~P., \& {Reynolds}, R.~T. 1993, \icarus, 101,
  108

\bibitem[{{Kobayashi} {et~al.}(2006){Kobayashi}, {Umeda}, {Nomoto}, {Tominaga},
  \& {Ohkubo}}]{kob06}
{Kobayashi}, C., {Umeda}, H., {Nomoto}, K., {Tominaga}, N., \& {Ohkubo}, T.
  2006, \apj, 653, 1145

\bibitem[{{London} {et~al.}(1981){London}, {McCray}, \& {Auer}}]{lon81}
{London}, R., {McCray}, R., \& {Auer}, L.~H. 1981, \apj, 243, 970

\bibitem[{{Lopez} \& {Fortney}(2013)}]{lop13}
{Lopez}, E.~D. \& {Fortney}, J.~J. 2013, \apj, 776, 2

\bibitem[{{Lyubarskii}(1990)}]{lyu90}
{Lyubarskii}, Y.~E. 1990, Soviet Astronomy Letters, 16, 16

\bibitem[{{Martin} {et~al.}(2016){Martin}, {Livio}, \& {Palaniswamy}}]{mar16}
{Martin}, R.~G., {Livio}, M., \& {Palaniswamy}, D. 2016, \apj, 832, 122

\bibitem[{{Nomoto} {et~al.}(2006){Nomoto}, {Tominaga}, {Umeda}, {Kobayashi}, \&
  {Maeda}}]{nom06}
{Nomoto}, K., {Tominaga}, N., {Umeda}, H., {Kobayashi}, C., \& {Maeda}, K.
  2006, Nuclear Physics A, 777, 424

\bibitem[{{Page} {et~al.}(2006){Page}, {Geppert}, \& {Weber}}]{pag06}
{Page}, D., {Geppert}, U., \& {Weber}, F. 2006, Nuclear Physics A, 777, 497

\bibitem[{{Patruno} \& {Watts}(2012)}]{pat12r}
{Patruno}, A. \& {Watts}, A.~L. 2012, ArXiv e-prints
  [\eprint[arXiv]{1206.2727}]

\bibitem[{{Pavlov} {et~al.}(2007){Pavlov}, {Kargaltsev}, {Garmire}, \&
  {Wolszczan}}]{pav07}
{Pavlov}, G.~G., {Kargaltsev}, O., {Garmire}, G.~P., \& {Wolszczan}, A. 2007,
  \apj, 664, 1072

\bibitem[{{Pfahl} \& {Rappaport}(2001)}]{pfa01}
{Pfahl}, E. \& {Rappaport}, S. 2001, \apj, 550, 172

\bibitem[{{Philippov} {et~al.}(2015){Philippov}, {Cerutti}, {Tchekhovskoy}, \&
  {Spitkovsky}}]{phi15}
{Philippov}, A.~A., {Cerutti}, B., {Tchekhovskoy}, A., \& {Spitkovsky}, A.
  2015, \apjl, 815, L19

\bibitem[{{Podsiadlowski}(1993)}]{pod93}
{Podsiadlowski}, P. 1993, in Astronomical Society of the Pacific Conference
  Series, Vol.~36, Planets Around Pulsars, ed. J.~A. {Phillips}, S.~E.
  {Thorsett}, \& S.~R. {Kulkarni}, 149--165

\bibitem[{{Rauscher} {et~al.}(2002){Rauscher}, {Heger}, {Hoffman}, \&
  {Woosley}}]{rau02}
{Rauscher}, T., {Heger}, A., {Hoffman}, R.~D., \& {Woosley}, S.~E. 2002, \apj,
  576, 323

\bibitem[{{Ruderman} {et~al.}(1989){Ruderman}, {Shaham}, \& {Tavani}}]{rud89}
{Ruderman}, M., {Shaham}, J., \& {Tavani}, M. 1989, \apj, 336, 507

\bibitem[{{Selsis} {et~al.}(2007){Selsis}, {Kasting}, {Levrard}, {Paillet},
  {Ribas}, \& {Delfosse}}]{sel07}
{Selsis}, F., {Kasting}, J.~F., {Levrard}, B., {et~al.} 2007, \aap, 476, 1373

\bibitem[{{Shannon} {et~al.}(2013){Shannon}, {Cordes}, {Metcalfe}, {Lazio},
  {Cognard}, {Desvignes}, {Janssen}, {Jessner}, {Kramer}, {Lazaridis},
  {Purver}, {Stappers}, \& {Theureau}}]{sha13}
{Shannon}, R.~M., {Cordes}, J.~M., {Metcalfe}, T.~S., {et~al.} 2013, \apj, 766,
  5

\bibitem[{{Sigurdsson} {et~al.}(2003){Sigurdsson}, {Richer}, {Hansen},
  {Stairs}, \& {Thorsett}}]{sig03}
{Sigurdsson}, S., {Richer}, H.~B., {Hansen}, B.~M., {Stairs}, I.~H., \&
  {Thorsett}, S.~E. 2003, Science, 301, 193

\bibitem[{{Smith} {et~al.}(2004){Smith}, {Scalo}, \& {Wheeler}}]{smi04}
{Smith}, D.~S., {Scalo}, J., \& {Wheeler}, J.~C. 2004, \icarus, 171, 229

\bibitem[{{Stevens} {et~al.}(1992){Stevens}, {Rees}, \&
  {Podsiadlowski}}]{ste92}
{Stevens}, I.~R., {Rees}, M.~J., \& {Podsiadlowski}, P. 1992, \mnras, 254, 19P

\bibitem[{{Tanvir} {et~al.}(2013){Tanvir}, {Levan}, {Fruchter}, {Hjorth},
  {Hounsell}, {Wiersema}, \& {Tunnicliffe}}]{Tanviretal2013}
{Tanvir}, N.~R., {Levan}, A.~J., {Fruchter}, A.~S., {et~al.} 2013, \nat, 500,
  547

\bibitem[{{Taylor}(1991)}]{tay91}
{Taylor}, Jr., J.~H. 1991, IEEE Proceedings, 79, 1054

\bibitem[{{Tchekhovskoy} {et~al.}(2016){Tchekhovskoy}, {Philippov}, \&
  {Spitkovsky}}]{tch16}
{Tchekhovskoy}, A., {Philippov}, A., \& {Spitkovsky}, A. 2016, \mnras, 457,
  3384

\bibitem[{{Thorsett} {et~al.}(1993){Thorsett}, {Arzoumanian}, \&
  {Taylor}}]{tho93}
{Thorsett}, S.~E., {Arzoumanian}, Z., \& {Taylor}, J.~H. 1993, \apjl, 412, L33

\bibitem[{{Toropina} {et~al.}(2012){Toropina}, {Romanova}, \&
  {Lovelace}}]{tor12}
{Toropina}, O.~D., {Romanova}, M.~M., \& {Lovelace}, R.~V.~E. 2012, \mnras,
  420, 810

\bibitem[{{Tsuruta}(1979)}]{tsu79}
{Tsuruta}, S. 1979, \physrep, 56, 237

\bibitem[{{Wang} {et~al.}(2006){Wang}, {Chakrabarty}, \& {Kaplan}}]{wan06}
{Wang}, Z., {Chakrabarty}, D., \& {Kaplan}, D.~L. 2006, \nat, 440, 772

\bibitem[{{Wolszczan}(1994)}]{wol94}
{Wolszczan}, A. 1994, Science, 264, 538

\bibitem[{{Wolszczan} \& {Frail}(1992)}]{wol92}
{Wolszczan}, A. \& {Frail}, D.~A. 1992, \nat, 355, 145

\bibitem[{{Wyatt}(2008)}]{wya08}
{Wyatt}, M.~C. 2008, \araa, 46, 339

\bibitem[{{Yan} {et~al.}(2013){Yan}, {Shen}, {Yuan}, {Wang}, {Rottmann}, \&
  {Alef}}]{yan13}
{Yan}, Z., {Shen}, Z.-Q., {Yuan}, J.-P., {et~al.} 2013, \mnras, 433, 162

\end{thebibliography}

\end{document}